\documentclass[a4paper,fleqn,usenatbib]{mnras}

\usepackage[T1]{fontenc}
\usepackage{ae,aecompl}

\usepackage{savesym}
\usepackage{graphicx}	
\usepackage{amsmath}	
\usepackage{pdflscape}
\usepackage{adjustbox}
\savesymbol{iint}
\usepackage{amssymb}
\usepackage{txfonts}
\restoresymbol{TXF}{iint}

\usepackage{multicol}        
\usepackage{bm}	
\usepackage{rotating}
\title{Dust and inclination corrected star-formation and interstellar medium scaling relations in nearby galaxies}
\author[B. A. Pastrav]{Bogdan A. Pastrav$^{1}$\thanks{E-mail:bapastrav@spacescience.ro}
\\
$^{1}$Cosmology and Astroparticle Physics Laboratory, Institute of Space Science, Atomistilor 409, 077125, Bucharest-Magurele, Romania\\
}
\date{Accepted 2023 December 13. Received 2023 December 5; in original form 2023 May 29.}

\pubyear{2023}

\begin{document}
\label{firstpage}
\pagerange{\pageref{firstpage}--\pageref{lastpage}}
\maketitle

\begin{abstract}

Following from our recent work, we present a detailed analysis of star-formation and interstellar medium (ISM) scaling relations, done on a representative sample
of nearby galaxies. H$\alpha$ images are analysed in order to derive the integrated galaxy luminosity, known as a more instantenous and accurate star-formation 
rate (SFR) tracer, and the required photometric and structural parameters. Dust and inclination corrected H$\alpha$ luminosities, SFRs and related quantities
are determined using a self-consistent method based on previous work prescriptions, which do not require the assumption of a dust attenuation 
curve and use of Balmer decrements (or other hydrogen recombination lines) to estimate the dust attenuation, with the advantage of determining dust opacities and
dust masses along the way. We investigate the extent to which dust and inclination effects bias the specific parameters of these relations, the scatter and degree of 
correlation, and which relations are fundamental or are just a consequence of others. Most of our results are consistent within errors with other similar studies,
while others come in opposition or are inconclusive. By comparing the B band optical and H$\alpha$ (star-forming) discs scalelengths, we found on average, the 
star-formation distribution to be more extended than the stellar continuum  emission one (the ratio being 1.10), this difference increasing with stellar mass. Similarly,
more massive galaxies have a more compact stellar emission surface density than the star-formation one (average ratio of 0.77). The method proposed can be applied in larger
scale studies of star-formation and ISM evolution, for normal low to intermediate redshift galaxies.
\end{abstract}
\begin{keywords}
   galaxies: star formation -- ISM: dust, extinction -- ISM: evolution -- galaxies: evolution --  galaxies: spiral -- galaxies: ISM
\end{keywords}

\maketitle

\section{Introduction}\label{sec:intro}

Dust and star-formation scaling relations are essential in studies of interstellar medium (ISM) evolution, in star-formation and galaxy evolution studies, or related to the
duty cycle of dust and gas in galaxies. Dust can be considered a good ISM tracer even though it is found in quantities of only up to $\simeq1\%$ of the total ISM mass 
(\citealt{DraineLee84, Draine03}), the rest being mostly gas in different phases and forms (atomic/neutral, molecular or ionized hydrogen). It is also a processor of stellar 
radiation as it scatters and absorbs the stellar radiation in ultraviolet (UV) and optical and emits it at longer wavelengths, in mid-infrared (MIR) to the far-infrared (FIR)
domain (\citealt{Gal18}). Besides being present in significant quantities in the discs of spiral galaxies (\citealt{Tuf02, Pop11, Vla05, Dri07, Dar11, Row12, Bou12,
Dal12} - the \textit{diffuse} dust distribution, it also surrounds the birthclouds of stars in the star-forming regions - the \textit{localized} distribution, obscuring the 
radiation coming from the young stars and is a nuissance in estimations of star-formation rates and of the fraction of radiation which escapes the birthclouds of stars into the
ISM (\citealt{Ken98, Noll09}).

Star-formation rates (SFR), which quantify the star-formation process - the transformation of cold gas into stars, are pivotal quantities in the attempt to understand and
characterise galaxy evolution. As galaxies can have various ISM properties and be in different stages of star-formation (e.g. actively star-forming, starbursts, quiescent),
deriving consistent and unbiased values for the star-formation rates based on different proxies or tracers is a real problem, and it produces important differences in the
values obtained (\citealt{Dal07, Noll09}). Another aspect to be considered when using different tracers to estimate SFR is that most of them are affected by various systematic
biases, such as dust attenuation, considering a constant initial mass function (IMF), the metallicity dependence, location within the galaxy and variations in the ISM conditions
(\citealt{KenEv12}). This in turn can significantly influence the related scaling relations and their characteristic parameters (e.g. slope, zero-point or correlation coefficient).\\
The most direct method of determination for star-formation rates is to count the number of stars of a certain age (Kennicutt \& Evans 2012), but at the level of current
instrumentation capabilities this method is limited to mostly Local Group galaxies. For more distant galaxies, the usual strategy to measure SFRs is to use UV continuum
(\citealt{Sal07}) and emission line tracers. Near-ultraviolet (NUV) continuum is one of the most direct tracers of recent star-formation as it traces the emission from young
stars (\citealt{KenEv12}). However, NUV observations are strongly affected by interstellar dust attenuation, this effect being less important at longer wavelengths, as shown
by \cite{Tuf04}, \cite{Mol06}, \cite{Gad10}, \cite{Pas12}, \cite{Pas13a} and \cite{Pas13b}. The UV slope (so called $\beta$) has been used to estimate the attenuation (\citealt{Hao11})
but this approach relies on many assumptions like the shape of the dust attenuation curve and dust geometry or the unknown intrinsic UV colors. 
Combination of FUV (far ultaviolet)+ infrared (IR) or TIR (total infrared) fluxes have been used by \cite{Ken09}, \cite{Hao11}, \cite{Ski11}, \cite{Whi14},
\cite{Remy15}, \cite{Hunt16}, \cite{Bar19}, \cite{Hunt19}, or FUV/NUV$+$MIR (e.g. $22\mu$m flux, \citealt{Ler21}), to derive star-formation rates corrected for dust
attenuation effects using an energy-balance approach (\citealt{Calz07, ZhuY08, Ken09}). Still, this method is also affected by dust attenuation,
which is often calculated from Balmer decrements, or by a stellar population age dependence (\citealt{KenEv12}). Moreover, the use of TIR and 24$\mu$m luminosities
has been shown to be problematic due to the contamination of dust heating by low mass older stars, as found in \cite{Ken09}, \cite{Boq14}, \cite{DeL14}, \cite{Via17}. \\
Another method is to use hydrogen recombination lines (with wavelengths in the optical range), such as the H$\alpha$ line flux/luminosity, in combination with other MIR fluxes as
H$\alpha$+8$\mu$m (\citealt{Calz07, Ken09}), H$\alpha$+24$\mu$m (\citealt{Ken07, Calz07, Ken09, Ski11, Remy15, Hunt16, Hunt19}), or even just the 24$\mu$m luminosity
(\citealt{Alo-Her06, Calz07, PiqLo16}. Other near-IR hydrogen recombination lines, such as the near-infrared Pa$\alpha$ and Pa$\beta$ lines, have been used to estimate SFR as 
these are far less affected by dust extinction, but at the same time are fainter for longer wavelengths and more sensitive to the density and temperature of the gas (\citealt{Calz13}).
These lines can probe higher optical depths than the Balmer decrements (\citealt{Liu13}), reveal more obscured star-formation regions than the former, as found by \cite{Tat15}, 
\cite{Cle22}, and have been used by \cite{Alo-Her06}, \cite{Calz07}, \cite{Tat15}, \cite{PiqLo16}, \cite{Gime22} or \cite{Cle22} to calibrate SFR indicators in the MIR to NIR. 

In this third paper of the series, following from Pastrav (2020) (where we focused on dust effects on disc scaling relations) and Pastrav (2021) (where the main 
bulge and early-type galaxy scaling relations where analysed, together with black-hole scaling relations and criteria for bulge and galaxy classification), hereafter Paper I
and Paper II, we concentrate on the star-formation and dust/ISM scaling relations, investigating the extent to which dust and inclination (projection) effects bias the 
specific parameters of these relations, such as slope, zero point, scatter and correlation coefficient, or produce underestimated values for the star-formation rates of 
galaxies and other associated parameters. Preliminary results have been shown in \cite{Pas23}. Through the method proposed, we try to reduce the uncertainties produced by dust attenuation in the measurements of relevant
quantities, especially in the SFR. We choose to use the $H\alpha$ optical emission line flux as a SFR tracer. We make use of H$\alpha$ galaxy images for the purpose of this
work, and, as in Papers I and II, we decompose each galaxy into its main components (bulge+disc). Then, we use the method of \cite{Pas13a} and \cite{Pas13b} and their 
numerical corrections for projection (inclination), dust and decomposition effects, to recover the needed corrected photometric and structural parameters involved in the
analysed scaling relations. The numerical corrections were derived by analysing and fitting simulated images of galaxies produced by means of radiative transfer calculations
and the model of \cite{Pop11}. The empirical relation found by \cite{Gro13} is used again here, tailored for the H$\alpha$ line wavelength to determine the new values
for the central face-on dust opacity ($\tau_{H\alpha}$), a parameter which is essential when applying the corrections for dust effects. When determining corrected H$\alpha$ 
luminosities and star-formation rates, our proposed method circumvents the need of assuming of a dust attenuation curve (usually a Galactic extinction curve or other similar
ones, as in \citealt{Calz07, Ken09, Mous10, Calz10, Hao11, Gime22, Pes21, Pes22}, etc.) and the use of Balmer decrements (\citealt{Kew02, Bri04, Mous06}) or
other hydrogen recombination lines to estimate the dust attenuation (assuming a foreground dust screen aproximation), which have been shown to be affected by various biases
or being inconsistent for different types of galaxies (\citealt{Mous06}). We derive SFR using the unattenuated H$\alpha$ luminosities to obtain more accurate and
instantenous star-formation rate values than would be derived through other methods. For most of the corrected relations we investigate the degree of correlation between
the parameters, calculate the scatter of these relations and analyse the implications of the main results for star-formation and galaxy evolution. We then discuss these 
results and compare with other relavant studies on nearby galaxies. Due to the self-consistent treatment of dust attenuation, the method proposed here significantly 
reduces the specific dust and inclination biases which plague the derivation of SFR and related quantities. The method can also be applied successfully in future larger scale
studies of star-formation and ISM evolution, at low to intermediate redshifts, as spectroscopic surveys of large samples of local and distant galaxies, narrow-band
emission-line imaging surveys, and large imaging surveys of nearby galaxies have or will become available, e.g. JPAS (Javalambre-Physics of the Accelerating Universe 
Astrophysical Survey, \citealt{Ben14}), J-PLUS (Javalambre-Photometric Local Universe Survey, \citealt{Cen19}) - the J0660 (6614\AA{}) filter, \& S-PLUS (Southern Photometric
Local Universe Survey, \citealt{MenOliv19}) - the J0660 (6614\AA{}) filter, MUSE (Multi-Unit Spectroscopic Explorer, \citealt{Bac10}) at VLT (Very Large Telescope), and others. \\
Our study comes to emphasize the importance of having accurate, unbiased derived star-formation rates and scaling relations in studies of ISM evolution and star-formation.

The paper is organised as follows. In Sect.~\ref{sec:sample} we present the galaxy sample used in this study, while in Sect.~\ref{sec:method} we describe the method used
for this analysis and the motivation for our choices. In Sect.~\ref{sec:results} we present the main results - the dust and inclination corrected star-formation and interstellar
medium scaling relations, together with all their characteristic parameters, and comment upon them in relation with other relavant studies in the literature. In Sect.~\ref{sec:discussion}
we discuss upon the possible sources of errors, differences with other studies, and the limitations of the method, while in Sect.~\ref{sec:conclusions} we summarise the results 
obtained in this study and draw conclusions.

\section{Sample}\label{sec:sample}

Our sample consists of 19 low-redshift spiral galaxies and 5 lenticulars, included in the SINGS (\textit{Spitzer} Infrared Nearby Galaxies Survey; \citealt{Ken03}) survey and
the KINGFISH project (Key Insights on Nearby Galaxies: a Far-Infrared Survey with \textit{Herschel}; \citealt{Ken11}). The galaxies were already analysed in B band in Paper I 
and Paper II, while another galaxy  - NGC 5194 (M51) was added here. For the purpose of this study, we needed the H$\alpha$ line images for the same galaxies (analysed previoulsy
in B band), which we extracted from the NASA/IPAC Infrared Science Archive (IRSA) and NASA IPAC Extragalactic Database (NED). As before, we exclude barred, dwarf and 
irregular galaxies from the KINGFISH sample, because we want to observe dust-free scaling relations, and at this point we cannot properly account for the effects of dust on the 
photometric and structural parameters of the former (barred galaxies), or for the more peculiar geometry of the latter (dwarfs and irregulars). Ellipticals from the KINGFISH 
survey are also not considered here as we focus here on star-formation and dust/ISM relations, for which studying late-type galaxies and lenticulars is more relevant for this
purpose. Most of the images were taken with the KPNO 1.5m (Kitt Peak National Observatory, t2ka detector - $0.304^{\prime\prime}$/pixel resolution) and CTIO
1.5m (Cerro Tololo Inter-American Observatory, $0.4344^{\prime\prime}$/pixel resolution) telescopes (see \citealt{Ken03, Ken09}), therefore having 
different sizes, resolutions or exposure times. For NGC 5033 no suitable H$\alpha$ image was found, therefore this galaxy was excluded in this study. \\
The KINGFISH project is an imaging and spectroscopic survey, consisting of 61 nearby (d<30 Mpc) galaxies, chosen to cover a wide range of galaxy properties (morphologies,
luminosities, SFR, etc.) and local ISM environments typical for the nearby universe, being therefore representative for the population of typical low redshift galaxies.

\section{Method}\label{sec:method}

The method used in this third part of our study is in general similar to the one used in Papers I and II when it comes to the fitting procedure, the sky determination and
subtraction, the photometry (now all done for the H$\alpha$ images), while the relations for the derivation of the dust opacity and dust mass were adapted to the H$\alpha$
line wavelength. Therefore, for a more detailed description, we refer the reader to Paper I, where the whole procedure is presented in great detail. Here, we just resume 
the whole procedure into a more concise version, given below.

\subsection{Fitting procedure}\label{sec:fitting}

For the fitting procedure of the H$\alpha$ line images of the galaxies in our sample, just as for the B band images, we used the GALFIT (version 3.0.2) data analysis 
algorithm (\citealt{Peng02, Peng10}). GALFIT uses a non-linear least-squares fitting based on the Levenberg-Marquardt algorithm. For the structural analysis 
(bulge-disc decomposition) of each galaxy and to fit the observed surface brightness of the spirals and lenticulars, we used the exponential (``expdisc'') and the S\'{e}rsic 
(``sersic'') functions available in GALFIT, for the disc and bulge surface brightness profiles, while the "sky" function was used for an initial estimation of the background
in each image.

As in our previous works (Papers I and II), the free parameters of the fits are: the X and Y coordinates of the centre of the galaxy in pixels, the bulge and disc integrated magnitudes, the
disc scale-length / bulge effective radius (for exponential/S\'{e}rsic function), axis-ratios of discs and bulges, bulge S\'{e}rsic index (for S\'{e}rsic function), the 
sky background (only in the preliminary fit - Step 1, see Paper I) and the sky gradients in X and Y. The input values for the coordinates of galaxy centre were determined
after a careful inspection of each image. Initial values for the position angles (PA) and axis-ratios were taken from NED. Although the central coordinates are free parameters,
we imposed a constraint on the fitting procedure, ensuring that the bulge and disc components were centred on the same position. The axis-ratio is defined as the ratio 
between the semi-minor and semi-major axis of the model fit (for each component). The position angle is the angle between the semi-major axis and the Y axis (increasing 
counter clock-wise). To mask the pixels corresponding to the additional light coming from neighboring galaxies, stars, compact sources, AGN or image artifacts, for each 
galaxy image we used a complex star-masking routine to create a bad pixel mask. This was used as input in GALFIT.

\subsection{Sky determination and subtraction. Photometry}\label{sec:sky}

Following the procedure in three steps described in Paper I to estimate as accurate as possible the background level, we calculate the integrated fluxes for each galaxy,
together with the corresponding bulge-to-disc ratios and then derive all the structural and photometric parameters (this time at H$\alpha$ wavelength). The integrated (total) 
flux of each galaxy is calculated from the maximum curve-of-growth (CoG) value (in counts), at the $R_{max}$ galactocentric radius (this is defined as the radius beyond 
which there is no galaxy emission and, therefore, the CoG is basically flat towards larger radii). As for the B band images in Paper I, the uncertainties of the fluxes are
estimated from the root mean square of the CoG values from the first 10 elliptical anulli beyond $R_{max}$. 
The bulge-to-disc ratio ($B/D$) is estimated from the disc and bulge CoGs and compared with the one determined by the ratio of the total counts of the decomposed disc and
bulge images, as it has to be consistent, within errors. We have used again the positive sky residuals in the outer parts of galaxies (towards $R_{max}$ and beyond) to 
estimate the systematic errors in bulge-to-disc ratios. We determine here the H$\alpha$ bulge-to-disc ratios just to compare with the B band values (one would expect the 
former to be higher, because the bulge becomes more prominent at longer wavelengths) as these are not really necessary for the purpose of this study.\\
Here, in deriving the H$\alpha$ line fluxes, we have to take into consideration the contamination of the flux values by the $N[II]\lambda6548,6584$ lines, positioned very close in the
spectrum with respect with this Balmer line ($\lambda(H\alpha)=6563\AA{}$). Therefore, the initially derived values have to be corrected for the effect of this
line deblending / mixing, using the $N[II]/H\alpha$ ratios available in the literature. In this study, we have chosen to use the values derived in \cite{Ken09} (see their
Table 1), and multiplied the initially derived disc fluxes with a correction factor, $f_{corr}$, as in the equation below:
\begin{eqnarray}\label{eq:f_Halpha_corr}
 F_{d}^{obs}(H\alpha)=F_{d}^{obs}([NII]+H\alpha)\times f_{corr}  ,
\end{eqnarray}
with $f_{corr}=1/([NII]/H\alpha+1)$, and $F_{d}^{obs}([NII]+H\alpha)$ the contaminated H$\alpha$ disc flux. Then, we also corrected the new values for foreground extinction,
$A_{ext}$, derived by considering the values at optical wavelengths taken from NED, as in \cite{SF11} recalibration of the \cite{Schl98} infrared based dust map, 
and interpolating at the H$\alpha$ line wavelength. This gives approximately $A_{ext}(H\alpha)=0.6A_{ext}(B)$.
As the H$\alpha$ line emission is concentrated in the young stellar disc of galaxies, we use the integrated disc flux to further derive the observed (measured) H$\alpha$ luminosities
of the sample (avoiding any bulge flux contamination this way), according with the general formula
\begin{eqnarray}\label{eq:L_Halpha_obs}
 L^{obs}(H\alpha)=4\pi d_{gal}^{2}F_{d}^{obs}(H\alpha)
\end{eqnarray}
with $d_{gal}$ - the distance to each galaxy.
The derived integrated fluxes (in $erg/cm^{2}/s$) and the corresponding H$\alpha$ luminosities (both in log scale), the bulge-to-disc 
ratios for all galaxies of our sample are given in Table~\ref{tab:photo_fluxes}, together with the distances to each galaxy used in this study (the same as in {Papers I and II}), taken from NED.

\begin{table*}
\caption{\label{tab:photo_fluxes} The calculated fluxes for our sample (H$\alpha$ line), corrected for $N[II]\lambda6548,6584$ contamination and foreground extinction. The
columns represent: (1) - galaxy name; (2) - distance to each galaxy, in Mpc, taken from NASA Extragalactic Database (NED), as derived in: $a$ - \protect\cite{Tul13}, $b$ - \protect\cite{Kre17},
$c$ - \protect\cite{Dal09}, $d$ - \protect\cite{Jang12}, $e$ - \protect\cite{Man11}, $f$ -  \protect\cite{Poz09}, $g$ -  \protect\cite{Sor14}, $h$ -  \protect\cite{McQ16},
$i$ - \protect\cite{Theu07}, $j$ - \protect\cite{Tul88}, $k$ - \protect\cite{Sabbi18}; (3) bulge-to-disk ratios ($B/D$) derived from the decomposed images, with systematic
uncertainties derived as described in Sec.\ref{sec:sky}; (4) - the integrated flux for each galaxy, in $erg/cm^{2}/s$ (log scale); (5) - the error for the galaxy flux (log scale);
(6), (7) - the integrated fluxes of the disc component and the corresponding uncertainty, in $erg/cm^{2}/s$ (log scale).}
 \begin{tabular}{{r|r|r|r|r|r|r}}
  \hline \hline
    $Galaxy$   &  $d_{gal}$  &  $B/D$    &   $log(F_{gal}^{obs})$  &   $log(\sigma_{F_{gal}})$  &   $log(F_{d}^{obs})$ &     $log(\sigma_{F_{d}})$\\
               &   [Mpc]   &             &    $[\frac{erg}{cm^{2}s}]$   &   $[\frac{erg}{cm^{2}s}]$   & $[\frac{erg}{cm^{2}s}]$   &  $[\frac{erg}{cm^{2}s}]$ \\
                (1)  & (2)  &  (3)  &  (4) &  (5)  &  (6)  &  (7)\\
  \hline 
NGC 0024  &    $7.67^{a}$  &	   $0.00_{-0.00}^{+0.00}$    &    -11.50       &   -12.23    &    -11.50   &   -12.23\\  
NGC 0628  &    $9.59^{b}$  &       $0.05_{-0.00}^{+0.00}$    &    -10.99       &   -12.36    &    -11.01   &   -12.38\\   
NGC 2841  &   $14.60^{a}$  &       $0.21_{-0.03}^{+0.05}$    &    -10.70       &   -12.94    &    -10.78   &   -13.04\\  
NGC 2976  &   $ 3.57^{c}$  &       $0.00_{-0.00}^{+0.00}$    &    -10.85       &   -12.38    &    -10.85   &   -12.38\\  
NGC 3031  &   $ 3.62^{d}$  &       $1.26_{-0.05}^{+0.06}$    &    -10.79       &   -11.72    &    -11.15   &   -12.19\\         
NGC 3190  &   $24.20^{e}$  &       $0.36_{-0.08}^{+0.02}$    &    -12.34       &   -12.49    &    -12.47   &   -12.64\\  
NGC 3621  &   $ 6.73^{a}$  &       $0.03_{-0.01}^{+0.02}$    &    -11.88       &   -13.03    &    -11.89   &   -13.04\\  
NGC 3938  &   $17.90^{f}$  &       $0.03_{-0.00}^{+0.00}$    &    -11.87       &   -12.48    &    -11.88   &   -12.49\\  
NGC 4254  &   $14.40^{f}$  &       $0.08_{-0.00}^{+0.00}$    &    -11.29       &   -13.00    &    -11.32   &   -13.04\\  
NGC 4450  &   $15.20^{g}$  &       $0.29_{-0.08}^{+0.08}$    &    -11.61       &   -12.12    &    -11.72   &   -12.30\\  
NGC 4594  &   $ 9.55^{h}$  &       $4.71_{-0.05}^{+0.06}$    &    -10.98       &   -11.32    &    -11.74   &   -12.11\\  
NGC 4736  &   $ 4.59^{a}$  &       $1.23_{-0.02}^{+0.03}$    &    -10.90       &   -12.11    &    -11.25   &   -12.49\\  
NGC 4826  &   $ 5.50^{g}$  &       $0.77_{-0.01}^{+0.00}$    &    -11.30       &   -12.33    &    -11.57   &   -12.56\\  
NGC 5055  &   $ 8.20^{g}$  &       $0.21_{-0.00}^{+0.00}$    &    -11.58       &   -12.66    &    -11.66   &   -12.74\\  
NGC 5474  &   $ 6.98^{a}$  &       $0.16_{-0.02}^{+0.03}$    &    -10.78       &   -13.28    &    -11.84   &   -13.34\\      
NGC 7331  &   $13.90^{a}$  &       $0.66_{-0.02}^{+0.03}$    &    -10.99       &   -12.16    &    -11.21   &   -12.38\\  
NGC 7793  &   $ 3.70^{g}$  &       $0.01_{-0.00}^{+0.00}$    &    -10.85       &   -13.03    &    -10.86   &   -13.04\\
NGC 1377  &   $21.00^{i}$  &       $1.22_{-0.01}^{+0.02}$    &    -12.33       &   -12.84    &    -12.68   &   -13.19\\  
NGC 1482  &   $19.60^{j}$  &       $2.61_{-0.00}^{+0.00}$    &    -11.46       &   -12.45    &    -11.90   &   -13.04\\
NGC 1705  &   $5.22^{k,a}$ &       $0.67_{-0.00}^{+0.00}$    &    -11.60       &   -12.46    &    -11.78   &   -12.74\\  
NGC 3773  &   $17.00^{j}$  &       $0.29_{-0.08}^{+0.08}$    &    -12.75       &   -13.60    &    -12.86   &   -13.69\\ 
NGC 5866  &   $14.70^{a}$  &       $0.28_{-0.05}^{+0.06}$    &    -11.89       &   -12.78    &    -12.00   &   -12.86\\   
NGC 5194  &   $ 7.55^{k}$  &       $0.34_{-0.03}^{+0.04}$    &    -11.07       &   -12.02    &    -11.19   &   -12.19\\
\hline
 \end{tabular}
\end{table*}

\subsection{Deriving star-formation rates}

The H$\alpha$ luminosity, calculated from Eq.~\ref{eq:L_Halpha_obs}, is needed in this study to derive first the measured star-formation rates (SFR) for the analysed galaxies, as it is known
to be a SFR tracer (e.g. \citealt{KenEv12}). It is known that star-formation rates derived based on the $H\alpha$ line or other hydrogen recombination lines (which arise from
HII regions throughout the galaxy) are more accurate and give a more instantaneous value for the SFRs (tracing more recent star-formation, <10 Myr; \citealt{KenEv12})
than when using UV continuum or a combination of UV+MIR/FIR fluxes, or extract the rates from fitting the spectral energy distribution (SED) of a galaxy (which gives an 
estimation of star-formation over the last 100-500 Myr).\\
In the literature, as already summarised in Sec.~\ref{sec:intro}, there is a wide range of studies to estimate SFRs, using the either a combination of FUV$+$TIR or FUV/NUV$+$MIR
fluxes, or a combination of the H$\alpha$ line flux/luminosity and other MIR fluxes (see references in Sec.~\ref{sec:intro}).
Since we had thoroughly derived the fluxes and luminosities for our small sample, together with the self-consistent calculation of dust opacities which attenuate the
$H\alpha$ fluxes, we considered that deriving corrected SFRs based only on the unattenuated H$\alpha$ luminosities to be considerably accurate. Therefore we do not use other
UV or MIR/FIR/TIR fluxes / luminosities in combination with the H$\alpha$ luminosity. We further motivate our choice in the following section. \\
To determine the observed (attenuated) star-formation rates, we use the calibration from \cite{Ken98} and convert from a Salpeter (\citealt{Salp55}) to a Chabrier (\citealt{Cha03})
initial mass function (IMF), as in \cite{Gime22}, obtaining 
\begin{eqnarray}\label{eq:SFR_obs}
 SFR^{obs}=4.4\times10^{-42}L^{obs}(H\alpha)
\end{eqnarray}
We also determine the specific star-formation rates, $sSFR$, for the galaxies in our sample - the ratio between SFR and stellar mass - $sSFR^{obs}=SFR^{obs}/M_{*}$. For the
purpose of investigating certain star-formation related scaling relations and because we did not calculate the molecular gas surface densities, we instead derive the observed 
star-formation surface densities for our galaxies as 
\begin{eqnarray}\label{eq:sigma_SFR_obs}
\Sigma_{SFR}^{obs}=\frac{SFR^{obs}}{2\pi R_{eff,d}^{2}(H\alpha)}
\end{eqnarray}
with $R_{eff,d}(H\alpha)$ as the effective observed H$\alpha$ disc radius.

\subsection{Dust opacity and dust mass derivation}\label{sec:dust_opacity}

\begin{table*}
\caption{\label{tab:dust} Dust masses and dust opacities for the H$\alpha$ line, derived using Eqs.\ref{eq:Mdust_Halpha},\ref{eq:Grootes1} and \ref{eq:ustar}. 
The different columns represent: (1) - galaxy name; (2) - H$\alpha$ face-on dust optical depth; (3) - stellar mass surface
densities; (4) - corrected stellar mass surface densities; (5) - stellar masses taken from: $a$ - \protect\cite{Ken09}, $b$ - \protect\cite{Remy15}, 
$c$ - \protect\cite{Grossi15}, $d$ - \protect\cite{Kar18}, $e$ - \protect\cite{Ski11}; $f$ - \protect\cite{Euf17}; $g$ - \protect\cite{Hunt19}; $h$ - \protect\cite{Leh19}
(6) - dust masses; (7) - corrected dust masses; (8),(9)- neutral hydrogen (HI) masses and their errors, taken from \protect\cite{Remy15} and \protect\cite{Grossi15} ; (10)
- (14) - standard deviation for $\tau_{H\alpha}^{f}, \mu_{*}, M_{*}$, $M_{dust}$ and $M_{dust}^{i}$. In square brackets we have the units in which these quantities are 
given. All quantities except dust optical depth are given in decimal logarithm unit scale.}
 \begin{tabular}{{r|r|r|r|r|r|r|r|r|r|r|r|r|r}}
  \hline \hline
 $Galaxy$ &  $\tau_{H\alpha}^{f}$ &  $log(\mu_{*})$ & $log(\mu_{*}^{i})$ & $log(M_{*})$ & $log(M_{dust})$ & $log(M_{dust}^{i})$ & $log(m_{HI})$ & $\sigma_{log(m_{HI})}$ & $\sigma_{\tau_{H\alpha}^{f}}$  & $\sigma_{log(\mu_{*})}$ & $\sigma_{log(M_{*})}$ & $\sigma_{log(M_{dust})}$  & $\sigma_{log(M_{dust}^{i})}$\\
          &                 &  $[\frac{M_{\odot}}{kpc^{2}}]$ & $[\frac{M_{\odot}}{kpc^{2}}]$  &  $[M_{\odot}]$ & $[M_{\odot}]$  &  $[M_{\odot}]$  & $[M_{\odot}]$  & $[M_{\odot}]$ &  &  $[\frac{M_{\odot}}{kpc^{2}}]$  & $[M_{\odot}]$  & $[M_{\odot}]$  & $[M_{\odot}]$ \\
              (1)  & (2)  &  (3)  &  (4) &  (5)  &  (6)  &  (7)  &  (8)  &  (9)  &  (10)  &  (11)  & (12)  &  (13)  & (14)\\
  \hline 
NGC 0024 &  2.21  &  7.99  &  9.10  & $ 9.48^{d}$  &  6.81  &  5.70  &  9.07  &  0.07  &  0.41  &  0.07  &  0.07  &  0.09 &   0.14\\
NGC 0628 &  0.66  &  7.51  &  7.55  & $10.29^{b}$  &  7.56  &  7.53  &  9.57  &  0.07  &  0.10  &  0.06  &  0.06  &  0.07 &   0.07\\
NGC 2841 &  1.22  &  7.76  &  8.46  & $10.17^{e}$  &  7.47  &  6.76  &  9.94  &  0.07  &  0.19  &  0.06  &  0.06  &  0.07 &   0.07\\
NGC 2976 &  2.33  &  8.01  &  8.68  & $ 8.96^{a}$  &  6.29  &  5.61  &  8.10  &  0.07  &  0.47  &  0.08  &  0.07  &  0.11 &   0.16\\
NGC 3031 &  2.44  &  8.02  &  8.73  & $10.39^{h}$  &  7.72  &  7.02  &  8.88  &  0.07  &  0.26  &  0.04  &  0.04  &  0.05 &   0.05\\
NGC 3190 &  1.14  &  7.73  &  8.61  & $10.03^{e}$  &  7.33  &  6.45  &  8.63  &  0.16  &  0.18  &  0.06  &  0.06  &  0.07 &   0.08\\
NGC 3621 &  1.08  &  7.71  &  8.41  & $ 9.43^{e}$  &  6.72  &  6.02  &  9.84  &  0.07  &  0.17  &  0.06  &  0.06  &  0.07 &   0.09\\
NGC 3938 &  0.33  &  7.25  &  7.32  & $ 9.46^{a}$  &  6.70  &  6.62  &  9.90  &  0.07  &  0.05  &  0.06  &  0.06  &  0.07 &   0.07\\
NGC 4254 &  0.68  &  7.53  &  7.64  & $ 9.61^{e}$  &  6.88  &  6.77  &  9.58  &  0.07  &  0.11  &  0.06  &  0.06  &  0.07 &   0.07\\
NGC 4450 &  3.28  &  8.14  &  8.47  & $10.40^{h}$  &  7.75  &  7.41  &  8.61  &  0.07  &  0.85  &  0.10  &  0.10  &  0.11 &   0.11\\
NGC 4594 &  3.80  &  9.20  & 10.92  & $10.97^{c}$  &  8.44  &  5.60  &  8.41  &  0.07  &  0.69  &  0.07  &  0.07  &  0.08 &   0.19\\
NGC 4736 &  3.80  &  8.76  &  9.12  & $10.21^{g}$  &  7.63  &  6.64  &  8.61  &  0.07  &  0.62  &  0.06  &  0.06  &  0.08 &   0.09\\
NGC 4826 &  1.97  &  7.94  &  8.32  & $ 9.99^{e}$  &  7.31  &  6.93  &  8.44  &  0.07  &  0.61  &  0.12  &  0.12  &  0.14 &   0.14\\
NGC 5055 &  2.27  &  8.00  &  8.87  & $10.49^{g}$  &  7.82  &  6.94  &  9.75  &  0.07  &  0.35  &  0.06  &  0.06  &  0.07 &   0.07\\
NGC 5474 &  0.55  &  7.44  &  7.52  & $ 9.06^{c}$  &  6.32  &  6.25  &  8.99  &  0.11  &  0.07  &  0.05  &  0.05  &  0.07 &   0.07\\
NGC 7331 &  1.22  &  7.76  &  8.58  & $10.56^{a}$  &  7.86  &  7.04  &  9.95  &  0.07  &  0.19  &  0.06  &  0.06  &  0.07 &   0.07\\
NGC 7793 &  2.04  &  7.96  &  8.25  & $ 9.47^{c}$  &  6.79  &  6.50  &  8.94  &  0.07  &  0.33  &  0.06  &  0.06  &  0.08 &   0.08\\
NGC 1377 &  2.15  &  7.98  &  8.46  & $ 9.28^{e}$  &  6.61  &  6.13  &  0.00  &  0.00  &  1.17  &  0.21  &  0.21  &  0.24 &   0.25\\
NGC 1482 &  0.24  &  7.13  &  7.58  & $ 9.99^{e}$  &  7.22  &  6.77  &  8.83  &  0.20  &  0.07  &  0.11  &  0.11  &  0.12 &   0.12\\
NGC 1705 &  0.62  &  7.49  &  7.48  & $ 8.19^{c}$  &  5.46  &  5.48  &  7.88  &  0.06  &  0.10  &  0.06  &  0.04  &  0.11 &   0.11\\
NGC 3773 &  0.49  &  7.40  &  7.62  & $ 8.31^{e}$  &  5.57  &  5.35  &  7.95  &  0.07  &  0.21  &  0.16  &  0.16  &  0.20 &   0.20\\
NGC 5866 &  2.27  &  8.00  &  8.89  & $10.02^{e}$  &  7.35  &  6.46  &  8.45  &  0.07  &  0.53  &  0.09  &  0.09  &  0.10 &   0.11\\
NGC 5194 &  2.74  &  8.07  &  8.68  & $10.53^{f}$  &  7.87  &  7.26  &  9.71  &  0.06  &  0.85  &  0.12  &  0.12  &  0.14 &   0.14\\
\hline
 \end{tabular}
\end{table*}

In Papers I and II (see section 3.3), we described the procedure and equations used to derive the dust opacity and dust mass for the analysed sample of spiral galaxies, in the
B band. Now we have to adapt the same equations for this study, at the H$\alpha$ line wavelength of 6563\AA{}. This is because the dust optical depth in the disc and the
associated dust mass will be different at this wavelength. 
Starting with Eq.(2) from \cite{Gro13} (but see also its derivation in Eqs.(A1-A5) from Appendix A of the same paper), that relates the dust mass at a wavelength
$\lambda$ to the corresponding central face-on dust opacity, we rewrite the aforementioned relation for the H$\alpha$ case:
\begin{eqnarray}\label{eq:Mdust_Halpha}
\tau_{H\alpha}^{f}=K(H\alpha)\frac{M_{dust}(H\alpha)}{R_{s,d}^{2}(H\alpha)}
\end{eqnarray}
This relation was calculated considering the dust geometry of the \cite{Pop11} model, where the diffuse dust in the disk (which mostly determines the optical depth of a 
spiral galaxy) is distributed axisymetrically in two exponential disks (see also Eq. (44) in \citealt{Pop11}). Therefore, the optical depth at a given wavelength
($\tau_{\lambda}$) and position will depend on the central face-on density of dust, e.g. the face-on opacity at a reference wavelength, $\lambda$. In Eq.~\ref{eq:Mdust_Halpha},
$K(H\alpha)$ is a constant containing the details of the dust geometry and the spectral emissivity of the \cite{Wei01} model. We had to recalculate it using the \cite{Pop11}
model equations (and model parameters from their E.1 table) and the dust model of \cite{Draine03}, now having a value of $0.6004 pc^{2}/kg$, considerably different than the 
value of $1.0089pc^{2}/kg$ found for the B band. This new value was obtained by interpolating the dust spectral emissivity values from the \cite{Draine03} model, $\kappa_{\lambda}$,
at the $H\alpha$ line wavelength, and deriving a new value of the two model dust exponential discs scalelength ratio, again at $\lambda_{H\alpha}$ (in the model it was 1.406
for the B band, as in their table E.1; we derived a value of 1.309 for our case).
Correspondingly, $R_{s,d}(H\alpha)$ is the scalelength of the H$\alpha$ stellar disc, in kpc.\\
Now, looking at the empirical correlation between $\tau_{B}^{f}$ and stellar mass surface density ($\mu_{*}$) of nearby spiral galaxies found by \cite{Gro13} that we used
in Papers I and II to derive the central face-on optical depth of the disc in the B band
\begin{eqnarray}\label{eq:Grootes}
\log(\tau_{B}^{f})=1.12(\pm0.11)\cdot\log(\mu_{*}/M_{\odot}kpc^{-2})-8.6(\pm0.8) ,
\end{eqnarray}
we had to evaluate the changes needed and their significance, for it to be valid for deriving $\tau_{H\alpha}^{f}$ - the dust opacity of the disc at H$\alpha$ line wavelength.
A more detailed discussion about this can be found in Appendix A. Here we just mention that for this case, we can rewrite Eq.~\ref{eq:Grootes} as
\begin{eqnarray}\label{eq:Grootes1}
\log(\tau_{H\alpha}^{f})=1.12(\pm0.11)\cdot\log(\mu_{*,H\alpha}/M_{\odot}kpc^{-2})-8.6(\pm0.8) ,
\end{eqnarray}
with $\mu_{*,H\alpha}$ being the stellar mass surface density (derived using the scalelength of the H$\alpha$ disc obtained through the bulge-disc decomposition)
\begin{eqnarray}\label{eq:ustar}
\mu_{*,H\alpha}=M_{*}/2\pi R_{s,d}^{2}(H\alpha)
\end{eqnarray}
The dust opacities, stellar mass surface densities and dust masses calculated using these relations are presented in Table~\ref{tab:dust}.

\begin{table}
\caption{\label{tab:struct_param} The photometric and structural parameters of the H$\alpha$ discs. The columns represent: (1) - galaxy name; (2) - the intrinsic disk axis-ratio, 
corrected for projection and dust effects; (3), (4) - the observed and intrinsic disk scalelengths; (5) - intrinsic bulge-to-disk ratio. In square
brackets we have the units in which these quantities are given.} 
\begin{tabular}{{r|r|r|r|r}}
 \hline \hline
 $Galaxy$  & $Q_{d}^{i}(H\alpha)$ & $R_{s,d}(H\alpha)$ & $R_{s,d}^{i}(H\alpha)$ & $(B/D)^{i}(H\alpha)$ \\
        &    &  [kpc]  &  [kpc]   &     \\
       (1)  & (2)  &  (3)  &  (4) &  (5) \\
   \hline 
NGC 0024   &     0.26 & 1.33 &  0.37 &   0.00\\ 
NGC 0628   &     0.95 & 5.23 &  5.58 &   0.06\\ 
NGC 2841   &     0.43 & 3.82 &  1.70 &   0.25\\ 
NGC 2976   &     0.50 & 0.71 &  0.33 &   0.00\\ 
NGC 3031   &     0.41 & 3.57 &  1.59 &   1.96\\ 
NGC 3190   &     0.34 & 3.36 &  1.22 &   0.40\\ 
NGC 3621   &     0.44 & 1.72 &  0.77 &   0.02\\ 
NGC 3938   &     0.91 & 3.04 &  2.78 &   0.02\\ 
NGC 4254   &     0.81 & 2.15 &  2.29 &   0.07\\ 
NGC 4450   &     0.70 & 2.96 &  2.18 &   0.52\\ 
NGC 4594   &     0.11 & 1.54 &  0.25 &   5.10\\ 
NGC 4736   &     0.69 & 1.26 &  0.83 &   1.40\\ 
NGC 4826   &     0.67 & 2.51 &  1.62 &   0.69\\ 
NGC 5055   &     0.41 & 3.80 &  1.53 &   0.18\\ 
NGC 5474   &     0.92 & 1.53 &  1.40 &   0.14\\ 
NGC 7331   &     0.38 & 6.00 &  2.33 &   0.69\\ 
NGC 7793   &     0.70 & 1.36 &  0.97 &   0.01\\ 
NGC 1377   &     0.59 & 1.07 &  0.61 &   1.25\\ 
NGC 1482   &     0.56 & 6.38 &  3.83 &   2.83\\ 
NGC 1705   &     0.93 & 0.53 &  0.54 &   0.88\\ 
NGC 3773   &     0.80 & 0.68 &  0.52 &   0.21\\ 
NGC 5866   &     0.38 & 1.62 &  0.88 &   0.21\\ 
NGC 5194   &     0.51 & 4.04 &  2.00 &   0.50\\ 
\hline
 \end{tabular}
\end{table}

\subsection{Correcting for dust, projection and decomposition effects}\label{sec:corr}

Once again, as in Papers I and II, in order to derive corrected values for all the parameters involved in the analysed dust/ISM and star-formation scaling relations, we used
the method developed and presented in \cite{Pas13a,Pas13b}. More specifically, we used the whole chain of corrections presented in Eqs.(4-13) from \cite{Pas13a} and 
Eqs.(3-13) from \cite{Pas13b}, together with all the numerical results (given in electronic form as data tables at CDS - Centre de Donn\'{e}es astronomiques de Strasbourg) to
correct the measured parameters for projection (inclination), dust and decomposition effects, in order to obtain their dust-free, intrinsic values. As now we analysed the images of
$H\alpha$ emission which comes from the young stellar disc of galaxies, we used the numerical corrections for the young stellar disc, already derived in \cite{Pas13a}
for the H$\alpha$ line. Then we proceeded similarly as in Paper I to correct all the necessary photometric and structural parameters (see Eqs. (4-9) \& (12-14) in Paper I for discs
and Eqs. (1-5) \& (8-12) in Paper II for bulges) for all the mentioned biases. The photometric parameters are also corrected for foreground extinction and cosmological
redshift dimming (the latter in the range of $0.01-0.05$ mag.). K-corrections or evolutionary ones were not applied as all the galaxies are at low redshift. \\
In the case of star-formation rates, the $H\alpha$ luminosity was the quantity that had to be debiased, in order to obtain corrected $SFR$ value for our sample. We can write the unattenuated luminosity as
\begin{eqnarray}\label{eq:L_Halpha_corr}
 L(H\alpha)^{corr}=L(H\alpha)^{obs}e^{\tau_{H\alpha}}=L(H\alpha)^{obs}10^{\frac{A_{H\alpha}}{2.5}} ,
\end{eqnarray}
with $A_{H\alpha}=1.086\tau_{H\alpha}$ being the attenuation of the H$\alpha$ line emission.
However, the dust opacity of the emission line is usually not equal with the dust opacity of the stellar continuum - the opacity of the starlight heating the dust, which
depends on many factors (dust attenuation curve, dust geometry, the SED of the stellar populations that heat the dust, etc.). But, as it has been shown in \cite{Ken09}, 
for the particular case of the H$\alpha$($\lambda6563\AA{}$) line, this aproximation holds, with the exception of more extreme cases. Considering this aproximation
valid for our case, we use the $\tau_{H\alpha}$ values derived from Eq.~\ref{eq:Grootes1} to account for the dust attenuation of our SFR tracer, the H$\alpha$ luminosity.
With this choice, we avoid the need of assuming of a dust attenuation curve (for example, a Galactic extinction curve or other similar ones) and the use of either the Balmer
decrements (the H$\alpha$/H$\beta$ ratios, \citealt{Calz00, Kew02, Bri04, Mous06, Pes21, Pes22}), or other near-IR hydrogen recombination lines and ratios between them, 
such as Paschen (Pa$\alpha$, Pa$\beta$,  eg. \citealt{PiqLo16}); Pa$\alpha$/H$\alpha$ \& Pa$\beta$/H$\alpha$ ratios (e.g. \citealt{Alo-Her06, Calz07, Liu13, Cle22, Gime22})
or Brackett lines ($Br\gamma$, \citealt{PiqLo16}), which may introduce systematic errors when deriving dust attenuations (with a greater extent for the Balmer line ratio). 
Our values for the attenuation of the emission line have been self-consistently derived, with a fixed star-dust geometry introduced in the calculation of dust opacities and 
dust masses, which can also introduce some systematic errors. Still, as the relations in Eqs.~\ref{eq:Mdust_Halpha} and \ref{eq:Grootes1} have been calibrated on a 
representative large sample of low-redshift spiral galaxies, we choose to use the $\tau_{H\alpha}$ values to correct the $L(H\alpha)^{obs}$ luminosities (as
in Eq.~\ref{eq:L_Halpha_corr}) instead of combining it with an additional MIR/FIR line (as an observational dust attenuation proxy). We further use $\tau_{H\alpha}$ to determine
the corrected (intrinsic) SFR, sSFR and $\Sigma_{SFR}$ as follows:
\begin{eqnarray}\label{eq:SFR_corr}
SFR^{corr}=4.4\times10^{-42}L(H\alpha)^{corr}=4.4\times10^{-42}L(H\alpha)^{obs}e^{\tau_{H\alpha}} 
\end{eqnarray}
\begin{eqnarray}\label{eq:sSFR_corr}
sSFR^{corr}=SFR^{corr}/M_{*} 
\end{eqnarray}
\begin{eqnarray}\label{eq:sigma_SFR_corr}
\Sigma_{SFR}^{corr}=SFR^{corr}/2\pi (R_{eff,d}^{i})^{2}(H\alpha)
\end{eqnarray}
with $R_{eff,d}^{i}$ being the intrinsic effective radius of the disc at H$\alpha$ wavelength.
The observed and intrinsic photometric and structural parameters needed for this study are shown in Table~\ref{tab:struct_param}. Likewise, all the star-formation
related parameters - H$\alpha$ luminosities, SFR, sSFR, SFR surface densities and their corresponding uncertainties (derived as described in the following section) are displayed
in Table~\ref{tab:SFR}.

\begin{table*}
\caption{\label{tab:SFR} The star-formation rates and the rest of related parameters (calculated using Eqs.~\ref{eq:L_Halpha_obs}-\ref{eq:sigma_SFR_obs} \& Eqs.~(\ref{eq:L_Halpha_corr}-\ref{eq:sigma_SFR_corr})
and their uncertainties. The columns represent: (1) - galaxy name; (2)-(4) - the H$\alpha$ observed and corrected luminosities, and the standard
deviations for $L(H\alpha)^{corr}$ (in decimal logarithm unit scale); (5)-(7) -
the observed and corrected star-formation rates, and standard deviations for $SFR$; (8)-(10) - the observed and corrected specific star-formation rates (in log scale), and
the corresponding standard deviations for $sSFR$; (11)-(13) - the observed and corrected SFR
surface densities (in log scale), and standard deviations for $\Sigma_{SFR}^{obs}$.}
 \begin{tabular}{{r|r|r|r|r|r|r|r|r|r|r|r|r}}
 \hline \hline
 \rotatebox{90}{$Galaxy$}  &  \rotatebox{90}{$log(L(H\alpha)^{obs})$} & \rotatebox{90}{$log(L(H\alpha)^{corr})$} & \rotatebox{90}{$log(\sigma_{L(H\alpha)^{corr}})$} & \rotatebox{90}{$SFR^{obs}$} & \rotatebox{90}{$SFR^{corr}$} & \rotatebox{90}{$\sigma_{SFR^{corr}}$} & \rotatebox{90}{$log(sSFR^{obs})$} & \rotatebox{90}{$log(sSFR^{corr})$} & \rotatebox{90}{$log(\sigma_{sSFR^{corr}})$} & \rotatebox{90}{$log(\Sigma_{SFR}^{obs})$} & \rotatebox{90}{$log(\Sigma_{SFR}^{corr})$} & \rotatebox{90}{$log(\sigma_{\Sigma_{SFR}^{corr}})$}\\
         \\
         &  $[\frac{erg}{s}]$ & $[\frac{erg}{s}]$ & $[\frac{erg}{s}]$ &  $[\frac{M_{\odot}}{yr}]$  & $[\frac{M_{\odot}}{yr}]$ & $[\frac{M_{\odot}}{yr}]$ & $[\frac{1}{yr}]$ & $[\frac{1}{yr}]$ & $[\frac{1}{yr}]$ & $[\frac{M_{\odot}}{yr*kpc^{2}}]$  & $[\frac{M_{\odot}}{yr*kpc^{2}}]$ & $[\frac{M_{\odot}}{yr*kpc^{2}}]$\\
     \\
    (1)  & (2)  &  (3)  &  (4) &  (5)  &  (6)  &  (7)  &  (8)  &  (9)  &  (10)  &  (11)  & (12)  &  (13)   \\
  \hline
 NGC 0024 &    40.35 &  41.31  &  40.73  &   0.10  &   0.89  &   0.24 &  -10.49  &  -9.53 &  -10.11 &  -2.50 &  -0.43 &  -1.01\\
 NGC 0628 &    41.03 &  41.32  &  40.52  &   0.47  &   0.91  &   0.14 &  -10.62  & -10.33 &  -11.13 &  -3.01 &  -2.78 &  -3.36\\
 NGC 2841 &    41.62 &  42.15  &  41.35  &   1.86  &   6.28  &   0.98 &   -9.90  &  -9.37 &  -10.18 &  -2.14 &  -0.91 &  -1.72\\
 NGC 2976 &    40.33 &  41.34  &  40.66  &   0.09  &   0.96  &   0.20 &   -9.99  &  -8.98 &   -9.66 &  -1.98 &  -0.30 &  -0.98\\
 NGC 3031 &    40.04 &  41.10  &  40.27  &   0.05  &   0.56  &   0.08 &  -11.70  & -10.64 &  -11.48 &  -3.66 &  -1.90 &  -2.59\\
 NGC 3190 &    40.37 &  40.87  &  40.71  &   0.10  &   0.33  &   0.23 &  -11.01  & -10.52 &  -10.68 &  -3.28 &  -1.91 &  -2.07\\
 NGC 3621 &    39.84 &  40.31  &  39.55  &   0.03  &   0.09  &   0.02 &  -10.94  & -10.47 &  -11.24 &  -3.23 &  -2.06 &  -2.83\\
 NGC 3938 &    40.70 &  40.85  &  40.30  &   0.22  &   0.31  &   0.09 &  -10.11  &  -9.97 &  -10.51 &  -2.87 &  -2.64 &  -3.18\\
 NGC 4254 &    41.08 &  41.37  &  40.59  &   0.52  &   1.03  &   0.17 &   -9.89  &  -9.60 &  -10.38 &  -2.37 &  -1.95 &  -2.69\\
 NGC 4450 &    40.72 &  42.15  &  41.72  &   0.23  &   6.16  &   2.29 &  -11.03  &  -9.61 &  -10.04 &  -2.82 &  -1.13 &  -1.52\\
 NGC 4594 &    40.30 &  41.95  &  41.62  &   0.09  &   3.89  &   1.82 &  -12.03  & -10.38 &  -10.71 &  -2.68 &   0.55 &   0.22\\
 NGC 4736 &    40.15 &  41.80  &  41.04  &   0.06  &   2.77  &   0.48 &  -11.42  &  -9.77 &  -10.53 &  -2.66 &  -0.64 &  -1.40\\
 NGC 4826 &    39.98 &  40.84  &  40.35  &   0.04  &   0.30  &   0.10 &  -11.36  & -10.51 &  -10.99 &  -3.42 &  -2.18 &  -2.65\\
 NGC 5055 &    40.24 &  41.23  &  40.47  &   0.08  &   0.74  &   0.13 &  -11.61  & -10.62 &  -11.37 &  -3.52 &  -1.75 &  -2.48\\
 NGC 5474 &    39.92 &  40.16  &  39.35  &   0.04  &   0.06  &   0.01 &  -10.50  & -10.26 &  -11.06 &  -3.05 &  -2.74 &  -3.28\\
 NGC 7331 &    41.15 &  41.68  &  40.91  &   0.62  &   2.09  &   0.36 &  -10.77  & -10.24 &  -11.01 &  -3.01 &  -1.66 &  -2.36\\
 NGC 7793 &    40.36 &  41.24  &  40.47  &   0.10  &   0.77  &   0.13 &  -10.47  &  -9.58 &  -10.36 &  -2.51 &  -1.33 &  -2.06\\
 NGC 1377 &    40.05 &  40.98  &  40.77  &   0.05  &   0.42  &   0.26 &  -10.59  &  -9.66 &   -9.86 &  -2.62 &  -1.20 &  -1.40\\
 NGC 1482 &    40.76 &  40.86  &  40.34  &   0.25  &   0.32  &   0.10 &  -10.59  & -10.49 &  -11.00 &  -3.46 &  -2.91 &  -3.24\\
 NGC 1705 &    39.73 &  40.00  &  39.56  &   0.02  &   0.04  &   0.02 &   -9.81  &  -9.55 &   -9.98 &  -2.32 &  -2.07 &  -2.38\\
 NGC 3773 &    39.68 &  39.89  &  39.55  &   0.02  &   0.03  &   0.02 &   -9.99  &  -9.78 &  -10.12 &  -2.59 &  -2.15 &  -2.49\\
 NGC 5866 &    40.40 &  41.39  &  40.86  &   0.11  &   1.08  &   0.32 &  -10.97  &  -9.99 &  -10.52 &  -2.62 &  -1.10 &  -1.56\\
 NGC 5194 &    40.65 &  41.84  &  41.35  &   0.19  &   3.01  &   0.99 &  -11.24  & -10.05 &  -10.54 &  -3.17 &  -1.37 &  -1.81\\
 \hline
 \end{tabular} 
\end{table*}
\subsection{Error estimation}\label{sec:errors}

To estimate the systematic errors on the main photometric and structural parameters needed in this study, namely the ones that characterise the H$\alpha$ discs of spiral 
galaxies, we ran a new set of fits for a few galaxies. In this process, we fixed the sky value to the one found initially by GALFIT and added $\pm1\sigma$, or $\pm3\sigma$
($\sigma$ being the uncertainty in the sky level), leaving free the parameters of interest (mainly the $R_{s,d}$(H$\alpha$), the disc central surface brightness or the 
disc axis-ratio, $Q_{d}$), while all other parameters were also fixed to the values found by GALFIT. The systematic errors in the disk scalelengths and bulge effective 
radii were within the range 1-10 pixels (1-3 arcsecs). They were less significant for the axis-ratios, up to $0.01$. This approach of error estimation of bulge parameters
was also used before by \cite{Gao19} and \cite{Gao20}. The error over $d_{gal}$ (measured distance to the galaxy) was taken from NED. We then performed propagation of
errors in Eqs.~\ref{eq:L_Halpha_obs}-\ref{eq:sigma_SFR_obs} and Eqs.~(\ref{eq:Grootes1}-\ref{eq:sigma_SFR_corr}) to obtain the standard deviation ($\sigma$) for all the needed parameters.\\
Having estimated already the uncertainties of the H$\alpha$ integrated fluxes, we continued with the propagation of errors to calculate the standard deviations for $L(H\alpha)$, SFR and sSFR, for both the observed and corrected values.

\section{Results}\label{sec:results}
 
We show here the main results of this study - an analysis of the star formation (Sec.~\ref{subsec:SFR_rel}) and ISM scaling relations (Sec.~\ref{subsec:ISM_rel}).
Throughout this section, the analysed relations are plotted in the linear form $log(Y) - log(X)$. The best-fit for each relation (from a linear regression procedure) has
the general form $log(Y)=\beta+\alpha \times log(X)$, with $\alpha$ - the intercept and $\beta$ - the slope of the relation. Unless specified otherwise, all the intercepts
and slopes are given in the same units as the ones of $log(Y)$ and $log(Y)/log(X)$.
\begin{figure*}
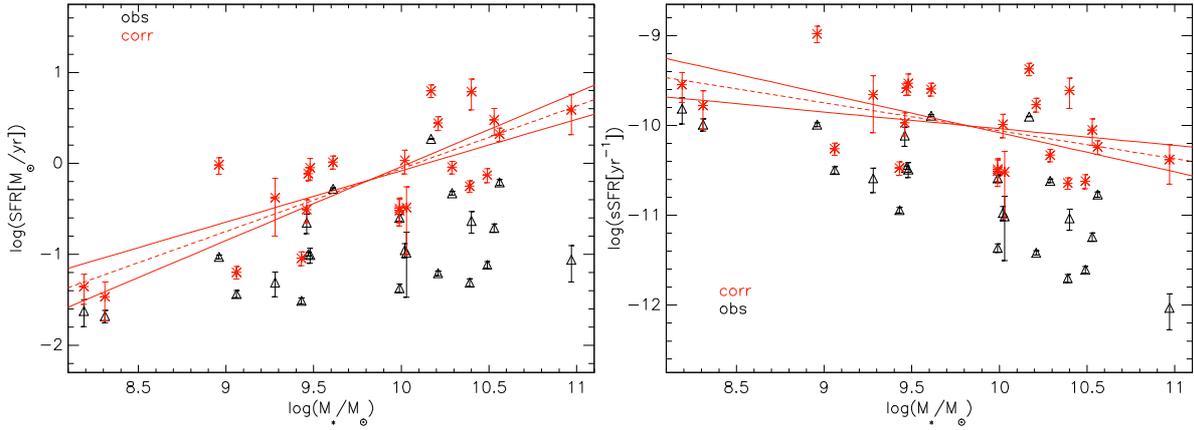

\begin{center}
 \includegraphics[scale=0.45]{SFR_vs_Mstar_Ha_v3.epsi}
 \includegraphics[scale=0.45]{sSFR_vs_Mstar_Ha_v1log.epsi}
\caption{\label{fig:SFR_MS} \textit{Left panel}: Star-formation main sequence, $SFR-M_{\ast}$, plotted in log scale. The observed SFR are shown with black triangles, while
the corrected rates are represented with red stars. The red dotted line is the SFR main sequence, obtained through a linear regression fit of the corrected values, 
while the two red solid ones delimit the $\pm1\sigma$ uncertainty range for the best-fit relation. The error bars represent the standard deviations.
\textit{Right panel}: Similar plot for the specific star-formation rate, $sSFR$, vs. stellar mass.}
\end{center}
\end{figure*}

\begin{figure*}
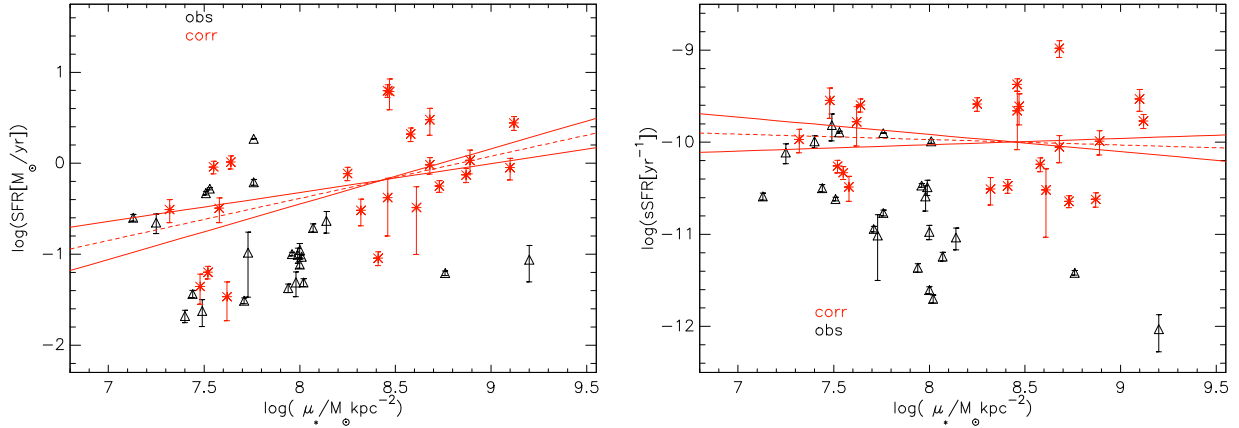

\begin{center}
 \includegraphics[scale=0.45]{SFR_vs_ustar_Ha_v1log.epsi}
 \hspace{0.33cm}
 \includegraphics[scale=0.45]{sSFR_vs_ustar_Ha_v1log.epsi}
\caption{\label{fig:SFR_MS_ustar} \textit{Left panel}: Star-formation rate versus stellar mass surface density, $SFR-\mu_{\ast}$, plotted in log scale. The observed SFR are
shown with black triangles, while the corrected rates are represented with red stars. The red dotted line is a linear regression fit of the corrected values, while
the two red solid ones delimit the $\pm1\sigma$ uncertainty range for the best-fit relation. The error bars represent the standard deviations. \textit{Right panel}: Similar
plot for the specific star-formation rate, $sSFR$.}
\end{center}
\end{figure*}

\subsection{Star-formation scaling relations}\label{subsec:SFR_rel}
The most important and well known star-formation relation is the one between $SFR$ and $M_{\ast}$, valid for local galaxies but also at higher redshifts (\citealt{Bri04, 
Noe07, Sal07, Elb11, Kar11, Whi12}), the "star-formation main sequence", SFMS. We show this relation in the left-hand panel of Fig.~\ref{fig:SFR_MS}. We recover the expected
trend, the linear increase of SFR with stellar mass. To obtain the specific parameters for the SFMS, we apply a linear reggresion fit for the corrected values and plot it as
a red dashed line, while the two red solid ones delimit the $\pm1\sigma$ uncertainty range for the best-fit relation. The zero-point, slope and scatter derived - $\beta=-6.95\pm1.22$, $\alpha=0.69\pm0.12$ and $\sigma=0.39$dex, are consistent and within errors with values 
calculated in other similar, larger scale studies. For example, \cite{Hunt16} found a value of 0.8 for a sample of galaxies from the local universe, including the KINGFISH 
galaxies, while \cite{Elb07} derived a value of 0.77, also for a sample of local galaxies. A higher slope of $0.89$ was determined by \cite{Gav13} for their HI-normal sample
of spiral galaxies taken mostly from the Virgo cluster. \cite{Whi12} determined a slope of $0.7$ for their low redshift sample, with a reduced degree of observed scatter of 0.34dex. A lower slope of $0.67$ and closer to our determined value was recently found by \cite{Cooke23}
for the low redshift (z=0.0-0.3) slice of his large sample of galaxies, selected to study the role of morphology and environment on the evolution of SFMS. While not plotted in
Fig.~\ref{fig:SFR_MS}, it is important to mention that the slope of the measured relation is severely lower than the one for the corrected SFMS, having a value of $\alpha=0.35\pm0.13$,
with a slightly increased scatter, $\sigma=0.41$dex. This underlines the importance of deriving dust and inclination corrected star-formation rates based on unbiased tracers.\\
In the right panel of Fig.~\ref{fig:SFR_MS}, we show a similar plot, this time with the relation between sSFR versus stellar mass. The decreasing trend with stellar mass is noticed,
as found in other studies (e.g. \citealt{Gav13, Grossi15, Hunt16}, etc.), with the slope of the corrected relation being this time shallower than for the observed one. We 
obtained $\beta=-6.95\pm1.22$, $\alpha=-0.31\pm0.12$ and $\sigma=0.40$dex in an analogus way as for the SFRMS. The slope for the corrected relation is in very
good agreement with the value of $-0.29$ found by \cite{Hunt16} for $z\sim0$ galaxies. \cite{Gav13} however, derived a significantly steeper relation, with a slope of $-0.56$ for 
their HI-normal sample. We do note here that the strength of the correlation for this relation is weaker than for the SFMS, with the derived correlation coefficient being 
$r_{sSFR,M_{\ast}}=-0.48$ (sSFR and $M_{\ast}$ are anticorrelated), as compared with the 0.77 value for the first relation. In a similar study, \cite{Gav13} found a value of
-0.46 for this correlation.\\
In Fig.~\ref{fig:SFR_MS_ustar} we show the SFR and sSFR of our sample, this time plotted as a function of stellar mass surface density. In this case, the trend in the corrected 
relations is maintained, however it is much shallower, with the corresponding slopes being $\alpha=0.46\pm0.14$ and $-0.05\pm0.12$, and with a higher degree of
scatter of 0.51 and 0.45dex. We can notice that there is practically no correlation between the sSFR and $\mu_{\ast}$, while the correlation $SFR-\mu_{\ast}$ is weaker
than that of SFMS, with the correlation coefficient being 0.57. To try and establish which of the relations is the fundamental one, the SFMS or the $SFR-\mu_{\ast}$ one, we used 
a linear partial correlation analysis of $(SFR, M_{\ast}, \mu_{\ast})$, by calculating the corresponding partial correlation coefficients. We found the following values:
$r_{SFRM_{\ast}, \mu_{\ast}}=0.656$, $r_{M_{\ast}\mu_{\ast}, SFR}=0.269$ and $r_{SFR\mu_{\ast}, M_{\ast}}=0.237$. From these values, one can say that indeed the stellar
mass is more important than the stellar mass surface density when deriving SFR, which suggests that the SFMS is the more fundamental relation.\\
\begin{figure}
 \includegraphics[scale=0.45]{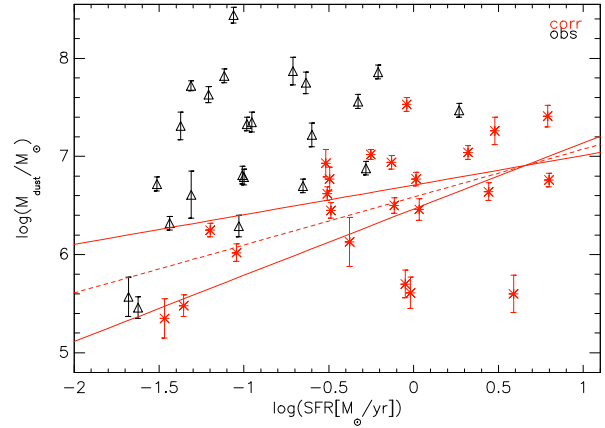}
  \caption{\label{fig:Md_SFR} The dust mass, $M_{dust}$, as a function of galaxy SFR. The symbols, colors and lines have the same meaning as those in Fig.~\ref{fig:SFR_MS}.} 
\end{figure} 
\begin{figure*}
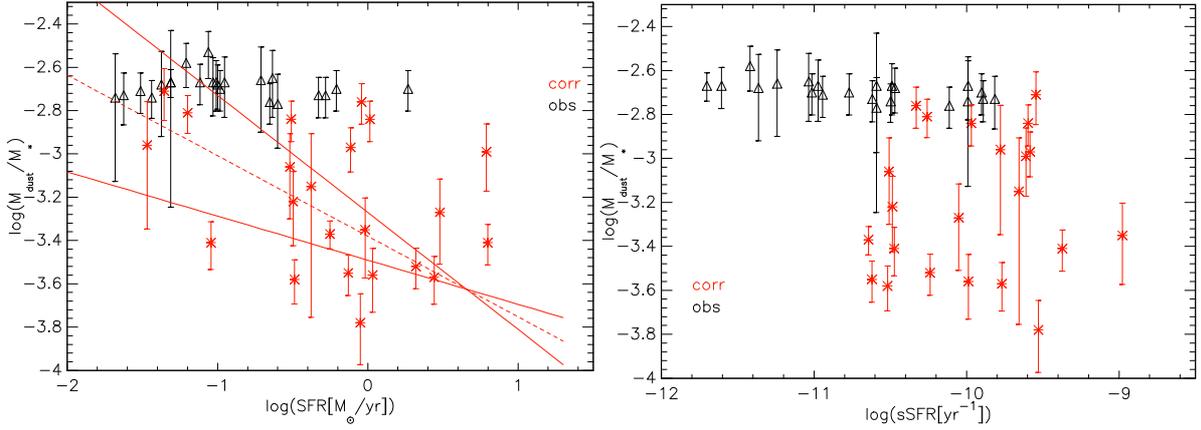

 \begin{center}
  \includegraphics[scale=0.45]{dust_to_star_vs_SFR_Ha_v3log.epsi}
 \includegraphics[scale=0.45]{dust_to_star_vs_sSFR_Ha_v2.epsi}
 \caption{\label{fig:DS_ratio_SFR} \textit{Left panel}: The dust-to-stellar mass ratios, $M_{dust}/M_{\ast}$, plotted against SFR. \textit{Right panel}: The same ratios, 
 plotted against the specific star-formation rates, sSFR. The symbols, lines and color legend are the same as in Fig.~\ref{fig:SFR_MS}. }
 \end{center}
\end{figure*}

Another relation that we present here is the one between dust mass, $M_{dust}$, and SFR, displayed in Fig.~\ref{fig:Md_SFR}. One can notice from the plot the increasing trend,
with more dust being found in galaxies with higher SFR. Considering the already tight relations $SFR - M_{\ast}$ and $M_{dust}-M_{\ast}$ (\citealt{Gro13,DeV17, Pas20, vanG22}), 
and the similar increasing behaviour observed, we might question if this relation is not in fact a consequence of the existence of these two relations. We try to answer this by
deriving the partial correlation coefficients for the quantities involved in these three relations, $(SFR, M_{\ast}, M_{dust})$. We find the following values:
$r_{SFRM_{\ast}, M_{dust}}=0.214$, $r_{M_{dust}M_{\ast}, SFR}=0.800$ and $r_{M_{dust}SFR, M_{\ast}}=0.286$. Keeping in mind that our sample is small, one can see from these
values that stellar mass is a more important quantity than SFR for the dust mass, making $M_{dust}-M_{\ast}$ the more important relation and the $M_{dust}-SFR$ relation a
secondary one. Of course, better statistics would be helpful here, considering the rather close values for $r_{SFRM_{\ast}, M_{dust}}$ and $r_{M_{dust}SFR, M_{\ast}}$.
A tight correlation between these quantities and similar increasing trend has been already observed by \cite{daC10} when analysing a large sample of SDSS galaxies.
They found a slope of $1.11\pm0.01$ and an intercept of $7.10\pm0.07$ for this relation. \cite{Hunt19}
have also explored this relation, finding a tight correlation with a scatter of 0.4-0.5 dex. We obtained for the corrected relation a slope of $\alpha=0.49\pm0.18$, with a
scatter of $\sigma=0.53$dex. The Pearson correlation coefficient calculated for the corrected relation confirms that this is a considerably tight one, having $r_{M_{d},SFR}=0.78$.
One can note here the higher degree of scattering in this relation than the one calculated for the SFMS relation. This could be due mostly to the few outliers that were not
excluded from the calculation of the coefficients, but also due to reduced size of our sample (better statistics from a larger sample would most likely reduce the scatter).
The similar parameters for the observed relation are considerably higher, with a slope of $\alpha=0.72\pm0.28$ and 0.64dex derived scatter.\\
Having analysed the SFMS, $M_{dust}-SFR$, and in Paper I the $M_{dust}-M_{\ast}$ relation, we further investigate here if there exists a correlation between the dust-to-stellar
mass ratio, $M_{dust}/M_{\ast}$, and SFR, as one might expect. In addition, we also plot this ratio as a function of sSFR. These plots are displayed in Fig.~\ref{fig:DS_ratio_SFR}. 
One can immediately notice that in the case of the observed ratios, there is a flat trend with SFR (as in the case of $M_{dust}/M_{\ast}-M_{\ast}$, see Paper I) and
sSFR, and therefore no correlation. Looking at the corrected dust-to-stellar mass ratios variation with the corrected SFR, it can be seen a slightly decreasing trend (slope -
$\alpha=-0.37\pm0.16$) of the dust-to-stellar mass ratios. This can be explained by the fact that more massive galaxies (and therefore older) with higher star-formation rates have
less dust available, as part of it has been destroyed by supernovae shocks / winds or other processes in the ISM. At the same time, the gas fraction decreases, less dust is
produced and the newly formed dust quantities can no longer overcome the destroyed mass of dust. The trend seen is expected considering the already observed decreasing behaviour
in the $M_{dust}/M_{\ast}-M_{\ast}$ (\citealt{Cort12, Grossi15, Pas20, Casa20}) relation, and the SFMS. This correlation is not so strong and tight, as we find
a low correlation coefficient, $r_{M_{dust}/M_{\ast},SFR}=-0.43$, with $\sigma=0.48$. In the second plot, there is no obvious increasing or decreasing trend with sSFR,
taking into consideration the associated large uncertainties. The downward trend is not obvious and the dependence of $M_{dust}/M_{\ast}$ on sSFR seems to be weak.
This result has also been previously found in \cite{Hunt19} for the KINGFISH galaxies, while \cite{Casa22} observed an apparently weak increasing behaviour for the resolved
version of this relation. The almost flat and inconclusive trend for $M_{dust}/M_{\ast}$ vs sSFR comes in opposition with the result found by \cite{Remy15} (see their Fig. 11),
\cite{Ski11}, and \cite{DeV17}, which show an increase in the dust-to-stellar ratio with sSFR for the KINGFISH sample. The same behaviour was observed by \cite{daC10} from
analysing a larger sample of low redshift SDSS galaxies.

One of the most important relations, derived from the SFMS relation shown in the left panel of Fig.~\ref{fig:SFR_MS}, is the one between the star-formation surface density, 
$\Sigma_{SFR}$, and the stellar mass durface density, $\mu_{\ast}$ ($\Sigma_{\ast}$ in other notations), previously named in the literature as the resolved star-formation main sequence
relation - rSFMS. This tight correlation was observed before in studies by \cite{San13}, \cite{C-D16}, \cite{G-D16}, \cite{Hsi17}, \cite{Med18}, \cite{ErF19}, \cite{Lin19},
\cite{Elli21}, \cite{Pes21} and \cite{Casa22}. In this study, we have not derived $\Sigma_{SFR}$ and $\mu_{\ast}$ for each valid spaxel of the galaxy images as in these studies,
but according with the formulas from Eqs.~\ref{eq:sigma_SFR_obs} and \ref{eq:sigma_SFR_corr}, this gives us a picture of the relation mostly at kpc-scale, and only for some
at sub-kpc scales. Nevertheless, a comparison of the characteristic parameters of this relation with those derived in previous works is still justified. We found the same
linearly increasing trend (in log scale) as in the already mentioned studies. Following a linear regression procedure, we found a slope for the corrected relation of $\alpha=1.03\pm0.18$,
a zero-point $\beta=-10.22\pm1.5$, with a rather large scatter of 0.44dex derived including all the galaxies, and larger than in the studies just mentioned here (e.g. 0.2-0.4dex).
The slope value is within the range of values found in these studies, e.g. 0.68 or 1.37 in \cite{Elli21}, 0.71 or 1.00 in \cite{Hsi17}, 0.88 in \cite{Casa22}, 1.19 in \cite{Lin19}, 1.04 in \cite{Pes21}
for example, depending on the linear regression method used, which may influence the final best-fit parameters, as shown in \cite{Hsi17}. The larger scatter in our study
could be due to our small sample, and therefore  inferior statistics, but the sample selection, the morphology and the variation of the global relation between galaxies
can also significantly contribute to this and influence the characteristics of the relation, as found by \cite{Elli21}. The Pearson correlation coefficient that
we calculated, $r_{\Sigma_{SFR},\mu_{\ast}}=0.79$, underlines the strength of this correlation. This value is considerably higher than those found by \cite{Lin19} (0.64),
\cite{Elli21} (0.57), for example, which may due to the fact that our best-fit relation is not so "local" as the ones derived in these studies. \cite{Casa22} found however 
a even higher correlation coefficient of 0.85 for their comparable small sample of nearby spiral galaxies, a part of them being present in our sample as well.
\begin{figure}
 \includegraphics[scale=0.45]{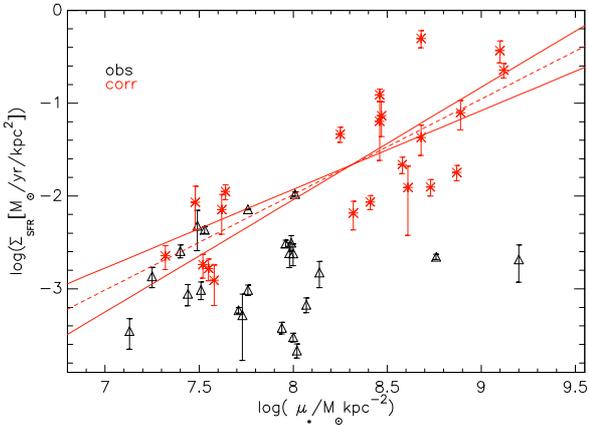}
 \caption{\label{fig:rSFMS} The resolved star-formation main sequence relation, rSFMS. The star-formation densities are derived according with Eqs.\ref{eq:sigma_SFR_obs}
and \ref{eq:sigma_SFR_corr}. The symbols, lines and color legend are the same as in Fig.~\ref{fig:SFR_MS}.}
\end{figure}

\begin{figure*}
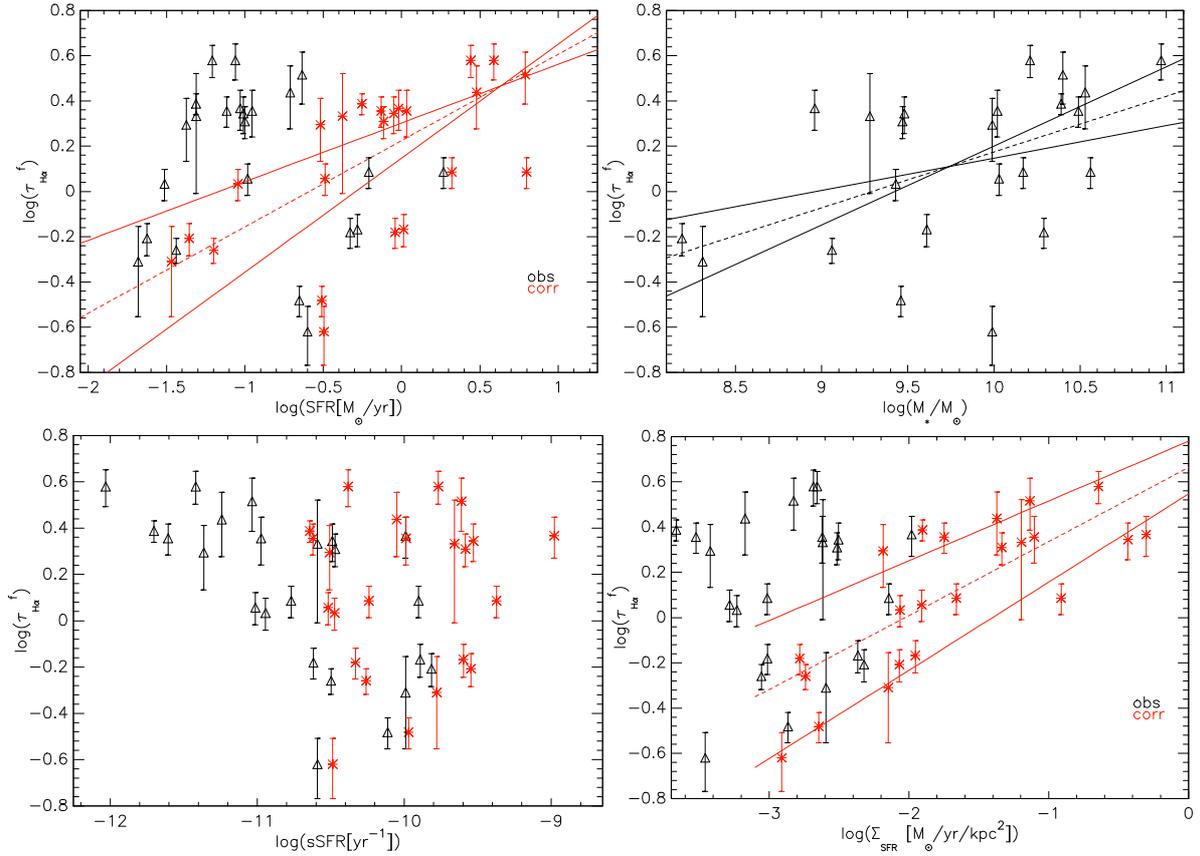

 \begin{center}
  \includegraphics[scale=0.45]{dust_opacity_vs_SFR_Ha_v1log.epsi}
 \hspace{-0.10cm}
 \includegraphics[scale=0.45]{dust_opacity_vs_Mstar_Ha_v1log.epsi}
 \vspace{0.10cm}
  \includegraphics[scale=0.45]{dust_opacity_vs_sSFR_Ha_v1log.epsi}
 \hspace{-0.10cm}
 \includegraphics[scale=0.45]{dust_opacity_vs_sigma_SFR_Ha_v1log.epsi}
 \caption{\label{fig:tau_SFR} The dust optical depth as a function of SFR (\textit{upper left panel}), $M_{\ast}$ (\textit{upper right panel}), sSFR (\textit{bottom left panel})
 and star-formation rate surface density, $\Sigma_{SFR}$ (\textit{bottom right panel}). The symbols, lines and color legend are the same as in Fig.~\ref{fig:SFR_MS}.}
 \end{center}
\end{figure*}
In Fig.~\ref{fig:tau_SFR} we check if there exists a relation between the dust central face-on optical depth and the star-formation rate, the stellar mass, specific star-formation rate, or the
star-formation rate surface density, $\Sigma_{SFR}$, given the already observed $M_{dust}-SFR$ relation in Fig.~\ref{fig:Md_SFR} and the direct dependence of $\tau_{H\alpha}^{f}$ 
on $M_{dust}$, given in Eq.~\ref{eq:Mdust_Halpha}. One can see an increase in dust opacity for galaxies with higher star-formation activity, and therefore higher dust production and
higher attenuation. This result was found in \cite{vanG22} too, but only for their low-redshift SDSS galaxies. Due to the aforementioned dependence, there is a correlation
between the dust opacities and star-formation rates, but weaker than in the case of $M_{dust}-SFR$ relation, with a correlation coefficient $r_{\tau_{H\alpha}^{f},SFR}=0.64$. 
The scatter and uncertainties in the vertical direction are larger than for the $M_{dust}-SFR$ relation, as dust opacity is a quantity more difficult to derive with great precision. 
However, it is rather difficult to establish which of the two relations - $M_{dust}-SFR$ or $\tau_{H\alpha}^{f}-SFR$ - is the fundamental one, due to the dependency
between the involved parameters, $(\tau_{H\alpha}, M_{dust}, SFR)$. In this respect, we did a linear partial correlation analysis for these quantities by deriving the corresponding
partial correlation coefficients. As a result, we found $r_{M_{dust}SFR, \tau_{H\alpha}}=0.959$, $r_{\tau_{H\alpha}M_{dust}, SFR}=-0.892$ and $r_{\tau_{H\alpha}SFR, M_{dust}}=0.937$. 
These would suggest that $M_{dust}$ dominates over $\tau_{H\alpha}$ when determining the SFR, and that $M_{dust}-SFR$ is the fundamental relation. However, as the values of the two
coefficients, $r_{M_{dust}SFR, \tau_{H\alpha}}$ and $r_{\tau_{H\alpha}SFR, M_{dust}}$ are very close, a more clear answer to this problem would require a larger sample. \\
In the upper right panel of Fig.~\ref{fig:tau_SFR}, we see that $\tau_{H\alpha}^{f}$ increases for galaxies with higher stellar masses, as also recently observed by \cite{vanG22} 
for a much larger sample of  galaxies at low and intermediate redshifts. It is worth mentioning that, if plotted as $\tau_{H\alpha}-log(M_{\ast}$), it is actually a curved 
relation, just as $\tau_{H\alpha}-log(SFR)$ and $\tau_{H\alpha}-log(\Sigma_{SFR})$ are too.
This relation is a result of the positive $M_{dust}-M_{\ast}$ correlation, observed by \cite{Gro13},
\cite{DeV17} and \cite{Pas20}. From the bottom left plot, we should note that no correlation between the dust opacity and corrected specific star-formation rate was found
($r_{\tau_{H\alpha},sSFR}$=-0.09), in opposition with the decreasing trend found in \cite{vanG22}, again only for their low redshift SDSS and GAMA galaxies. The anticorrelation
between dust optical depth and sSFR is observed for our measured relation only, which may lead us to the conclusion that dust and inclination effects or other biases 
have not been properly accounted for in the mentioned study. In the bottom right panel, we can also see an increase in dust opacity with the star-formation rate surface density
- a tracer for the molecular gas surface density (\citealt{Ler08, Schr11}), which fuels the star-formation. This slightly upwards trend was also found in 
\cite{vanG22} only for their low-redshift SDSS and GAMA samples, but not for the high redshift ones. It is important to note that this increasing trend is only seen in the corrected
relation, with a correlation coefficient $r_{\tau_{H\alpha},\Sigma_{SFR}}=0.79$, which shows the strength of this correlation, at least for low redshift galaxies. The existence
of this last relation reveals, at least for low redshift galaxies, the tight connection between star-formation fuel in the young stellar disc and dust mass distribution
(related directly with $\tau_{H_{\alpha}}^{f}$ as in Eq.~\ref{eq:Mdust_Halpha}) in the dust disc.
\begin{figure*}
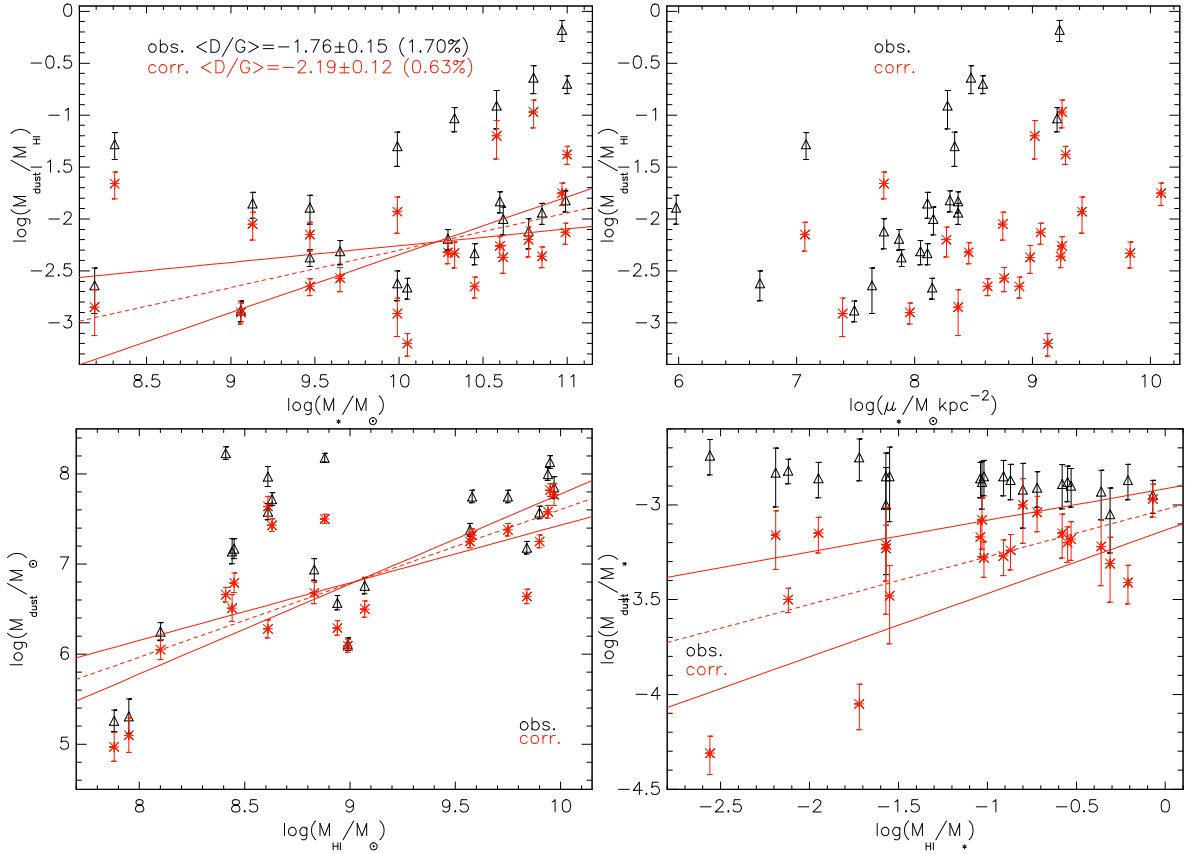

 \begin{center}
  \includegraphics[scale=0.45]{dust_to_HI_ratio_vs_stellar_mass_B_v3SpLen_log.epsi}
  \hspace{-0.10cm}
  \includegraphics[scale=0.45]{dust_to_HI_ratio_vs_stellar_mass_surf_density_B_v2SpLen.epsi}
  \vspace{0.10cm}
  \includegraphics[scale=0.45]{Md_vs_MHI_B_v2SpLen_log.epsi} 
   \hspace{-0.10cm}
  \includegraphics[scale=0.45]{Md-Mstar_vs_MHI-Mstar_B_v3SpLen_log.epsi}
  \caption{\label{fig:dust_gas_rel} Dust scaling relations. The four panels represent: the dust-to-HI (atomic hydrogen) mass ratio variation with galaxy stellar mass 
  (\textit{upper left}) and stellar mass surface density (\textit{upper right}); $M_{dust}$ vs. HI mass, $M_{HI}$ (\textit{bottom left}); the $M_{dust}/M_{\star}$ ratio vs. 
  HI-to-stellar mass ratio, $M_{HI}/M_{\star}$ (\textit{bottom right}). The symbols, lines and color legend are the same as in Fig.~\ref{fig:SFR_MS}.}
 \end{center}
\end{figure*}

\subsection{ISM scaling relations}\label{subsec:ISM_rel}
In the next figure, Fig.~\ref{fig:dust_gas_rel}, we show some of the dust scaling relations important for ISM studies, that can provide evidence about the role of dust in
the star-formation cycle and constrain chemical evolution models. On the upper row of plots, the dust-to-HI (atomic hydrogen) mass ratio variation with $M_{\ast}$
and $\mu_{\ast}$ are presented, $M_{HI}$ being the neutral hydrogen gas mass, taken from \cite{Remy15} and \cite{Grossi15}. On the bottow row, the $M_{dust}$ variation with
HI mass is shown, and the $M_{dust}/M_{\star}$ ratio vs. gas-to-star ratio, $M_{HI}/M_{\star}$ (bottom right). The increasing trends in the $M_{dust}/M_{HI}$ vs $M_{\star}$ and
$M_{dust}/M_{\star}$ vs $M_{HI}/M_{\star}$ are recovered after applying the corrections, and we find the average dust-to-gas ratio of our sample to be $-2.19\pm0.12$ (or $0.63\%$),
a value consistent with -2.1 one found by \cite{Cort12} for HI normal galaxies. This numerical ratio of $0.63\%$ is in line with the $\simeq1\%$ estimation of the dust mass
fraction in the ISM of galaxies. The increasing trend in $M_{dust}/M_{HI}$ towards more massive galaxies has also been noticed before in \cite{Cort12} (but shallower than in this
study), \cite{Grossi15} and \cite{DeV17}, and can be understood considering the relation between the stellar mass and gas metallicity (\citealt{Tre04}). The scatter for the
first relation is rather high - $\sigma=0.53$dex (compared with the 0.37dex scatter found by \citealt{Cort12}, for example) for the corrected one, while the corresponding 
correlation coefficient $r_{M_{dust}/M_{HI},M_{\star}}=0.32$ is consistent with the value of $r=0.31$ found in \cite{Cort12}, but slightly lower than the value derived by \cite{DeV17} (0.47). Practically 
no correlation is observed for the corrected second investigated relation, $M_{dust}/M_{HI}$ vs stellar mass surface density, $\mu_{\star}$, with $r_{M_{HI}/M_{\star},\mu_{\ast}}=0.06$,
which is in line with what was found by \cite{Cort12}, who derived a 0.10 value for this coefficient. \\
From the bottom left plot in Fig.~\ref{fig:dust_gas_rel}, we can see the strong correlation between dust and HI masses in the disc, for which we derive a correlation coefficient
$r_{M_{dust},M_{HI}}=0.94$ for the corrected relation, with a rather high degree of scatter, $\sigma=0.54$dex. Our derived coefficient is considerably higher than the ones found 
by \cite{DeV17} (0.74) or \cite{Casa20} (0.80, for the reversed relation, $M_{HI}$ vs. $M_{dust}$) by analysing larger samples of low-redshift spiral galaxies than ours. The 
slope found for this correlation, $\alpha=0.82\pm0.18$ is consistent with those found in previously mentioned studies - \cite{Casa20} found $0.85\pm0.03$ but for the reversed
relation. The tight correlation observed here should suggest that both dust and HI masses follow a similar radial distribution in the discs of galaxies. This is generally
the case for the dust, which is distributed in an exponential disc, while the HI distribution has a more complex form (\citealt{Casa17}).\\
A weaker correlation is found for the $M_{dust}/M_{\star}$ vs $M_{HI}/M_{\star}$ relation, with $r=0.36$, the increasing trend being also observed by \cite{Cort12}, \cite{DeV17} and
\cite{Casa20}, albeit more pronounced than in our plot (we derived a slope $\alpha=0.26\pm0.08$ for the corrected relation). For instance, \cite{DeV17} derived a slope of 0.47 and
found a very high correlation coefficient of 0.87, while \cite{Casa20} also found a much stronger correlation than in this study, with $r=0.80$, for a much larger sample of late-type
spiral galaxies, and a strong increasing trend with a slope of $1.21\pm0.05$, but for the reversed relation - $M_{HI}/M_{\star} - M_{dust}/M_{HI}$. As $M_{HI}/M_{\star}$ ratio is
considered an indicator of the evolutionary stage of a galaxy, and keeping in mind the $M_{dust}/M_{\star}-SFR$ variation seen earlier (and the explanation given for it), the fact
that we observe the existence of this last correlation in Fig.~\ref{fig:dust_gas_rel} means that $M_{dust}/M_{\star}$ is a measure of the evolutionary state of the galaxy. The
scatter of this last relation, $\sigma=0.32$dex is significantly more reduced than in the case of the $M_{dust}/M_{HI}$ vs $M_{\star}$ and $M_{dust}-M_{HI}$ relations. \\
While in other studies (as the just mentioned ones in this paragraph, and others) these dust scaling relations are studied as a function of environment (e.g. galaxies in 
clusters or groups vs field galaxies, HI normal vs. HI deficient galaxies), our sample is too reduced and formed out of galaxies with ISM environments characteristic for the 
nearby universe, to make such comparisons. Therefore we do not comment on these issues here.\\
A summary of the main results presented in this section, with all the derived parameters and the correlations found, for all the relations analysed, together with similar results found in the literature,
are shown in Table~\ref{tab:SFR_ISM_scal_relations}.

\begin{table*}
\begin{center}
 \caption{\label{tab:SFR_ISM_scal_relations} The linear regression best-fit parameters for all the star-formation and ISM corrected scaling relations presented,
 together with the correlation coefficients. In the last column, results from the literature are shown. The relations are in the form $log(Y)=\beta+\alpha \times log(X)$,
 with $\alpha$ - the intercept, $\beta$ - the slope, given in the same units as $log(Y)$ and $log(Y)/log(X)$. We also present here in the 4th and 5th columns, the scatter of the relations, $\sigma$
 (root mean square, in dex), and the Pearson correlation coefficient, $r$.}
 \begin{tabular}{{r|r|r|r|r|r}}
  \hline \hline
 Correlation      	&  $\beta$  &  $\alpha$  &  $\sigma$  &   $r$   &    Literature results; comments \\
 \hline
 $SFR-M_{\ast}$ \textbf{(SFMS)}     &  $-6.95\pm1.22$   & \textbf{0.69$\pm$0.12} &  0.39 &  \textbf{0.77}  &  \protect\cite{Hunt16}: $\alpha=0.80$;\\
                                    &                   &                        &       &                 &  \protect\cite{Elb07}: $\alpha=0.77$; \\
                                    &                   &                        &       &                 &  \protect\cite{Gav13}: $\alpha=0.89$; \\
                                    &                   &                        &       &                 &  \protect\cite{Whi12}: $\alpha=0.70, \sigma=0.34$ \\ 
                                    &                   &                        &       &                 &  \protect\cite{Cooke23}: $\alpha=0.67$; \\     
\hline
 $sSFR-M_{\ast}$     &  $-6.95\pm1.22$   & $-0.31\pm0.12$  &  0.40   &  -0.48    & 	 similar trend as in \protect\cite{Grossi15};\\
                     &                   &                        &       &                 &  \protect\cite{Hunt16}: $\alpha=-0.29$;\\
                     &                   &                 &         &           &       same trend and tight correlation in \protect\cite{daC10};\\
                      &                   &                 &         &           &                       \protect\cite{Gav13}:  $\alpha=-0.57, r=-0.46$ \\
 \hline  
 $SFR-\mu_{\ast}$  &  $-4.09\pm1.23$   & $0.46\pm0.14$ &  0.51 &  0.57  &  \\
 \hline          
 $sSFR-\mu_{\ast}$  &  $-9.49\pm1.08$   & $-0.05\pm0.12$ &  0.45 &  -0.10  & 	\textbf{no correlation!} \\
 \hline
 \textbf{$M_{dust}-SFR$}  &  $6.59\pm0.12$   & 0.49$\pm$0.18  &  0.53 &  \textbf{0.78}  &  \protect\cite{Hunt16}: $\sigma=0.40-0.50$; \\
                          &                  &                &       &                 &   \protect\cite{daC10}: $\alpha=1.11\pm0.01$ \\
 \hline
 $M_{dust}/M_{\ast}-SFR$  &  $-3.38\pm0.11$   & $-0.37\pm0.16$ &  0.48 &  -0.43  &  	\\
 \hline
 $M_{dust}/M_{\ast}-sSFR$  &  $-.--\pm-.--$   & $-.--\pm-.--$ &  0.52 &  0.21  &  	\textbf{no correlation!} (also in \protect\citealt{Hunt19}); \\
                      &                   &                 &         &           &   correlation found in \protect\cite{daC10,Ski11} \& \\
                      &                   &                 &         &           &          \protect\cite{Remy15,DeV17}  \\     
 \hline
 $\Sigma_{SFR}-\mu_{\ast}$  \textbf{(rSFMS)}  &  $-10.22\pm1.50$   & \textbf{1.03$\pm$0.18} &  0.44 &  \textbf{0.79}  &  \protect\cite{Hsi17}: $\alpha=0.71/1.00$;\\
                                      &                   &                        &       &                 &  \protect\cite{Lin19}: $\alpha=1.19, r=0.64$; \\
                                      &                   &                        &       &                 &  \protect\cite{Elli21}: $\alpha=0.68/1.37, r=0.57$; \\
                                      &                   &                        &       &                 &  \protect\cite{Pes21}: $\alpha=1.04\pm0.04, \sigma=0.44;$ \\
                                      &                   &                        &       &                 &  \protect\cite{Casa22}: $\alpha=0.88\pm0.03, r=0.85, \sigma=0.30$ \\
 \hline
 \textbf{$\tau-SFR$}     &  $0.22\pm0.07$   & 0.38$\pm$0.12 &  \textbf{0.27} &  \textbf{0.64}  &  same trend in \protect\cite{vanG22}(low-\it{z})		\\
 \hline
 $\tau-M_{\ast}$     &  $-2.29\pm1.00$   & $0.25\pm0.10$ &  0.30 &  0.64  &  same trend in \protect\cite{vanG22}		\\
 \hline
 $\tau-sSFR$     &  $-.--\pm-.--$   & $-.--\pm-.--$ & 0.34  &  0.09  &  	\textbf{no correlation!};\\ 
                     &                   &                 &         &           &     decreasing trend in \protect\cite{vanG22}(low-\it{z})\\			
 \hline
 \textbf{$\tau-\Sigma_{SFR}$}     &  $0.66\pm0.11$   & 0.33$\pm$0.06 &  \textbf{0.21} &  \textbf{0.79}  &  same trend in \protect\cite{vanG22}(low-\it{z})		\\
  \hline
 $M_{dust}/M_{HI}-M_{\ast}$     &  $-5.89\pm2.02$   & $0.36\pm0.19$ &  0.53 &  0.32  &  \protect\cite{Cort12}: $\sigma=0.37, r=0.31$;	\\
                                 &                   &                        &       &                 &  \protect\cite{DeV17}: $r=0.47$; \\
                             &                   &                 &         &           &  correlation found in \protect\cite{Grossi15} \\
 \hline                              
 $M_{dust}/M_{HI}-\mu_{\ast}$     &   $-.--\pm-.--$   & $-.--\pm-.--$ &  0.57 &  0.05  &  \textbf{no correlation!}; \protect\cite{Cort12}: $r=0.10$	 \\
 \hline
 \textbf{$M_{dust}-M_{HI}$}    &  $-0.59\pm1.62$   & \textbf{0.82$\pm$0.18} &  0.54 &  \textbf{0.94}  &  \protect\cite{Casa20}: $\alpha=0.85\pm0.03, r=0.80$; \\
                               &                   &                        &       &                 &  \protect\cite{DeV17}: $r=0.74$ \\
  \hline
 $M_{dust}/M_{\ast}-M_{HI}/M_{\ast}$     &  $-3.02\pm0.11$   & $0.25\pm0.08$ &  0.26 &  0.36  &  \protect\cite{DeV17}: $\alpha=0.47, r=0.87$;	\\
                                          &                   &                 &         &           &  correlation found in \protect\cite{Cort12} \\
 \hline               
 \end{tabular}
 \end{center}
\end{table*}

\subsection{The SF spatial distribution vs. optical stellar continuum emission extent}
\begin{figure}
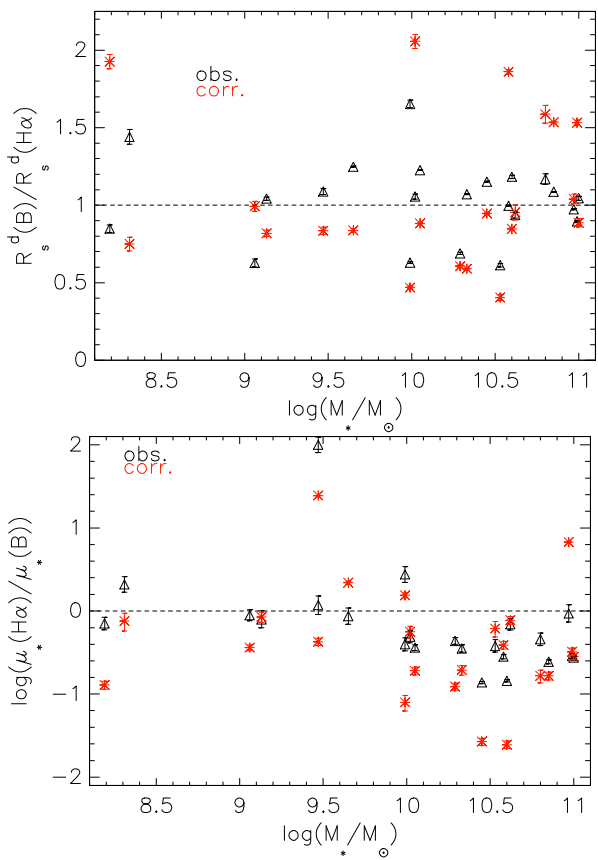

  \includegraphics[scale=0.45]{scalelength_ratio_HaB_vs_Mstar_v2SpLen.epsi}
    \includegraphics[scale=0.45]{ustar_ratio_vs_Mstar_Ha_vSpLen.epsi}
  \caption{\label{fig:Rs_ratio_Mstar} The ratio of the intrinsic disc scale-lengths seen in optical B band and in $H\alpha$ line as a function of stellar mass (\textit{upper plot}).
  The stellar mass surface density ratio vs $M_{\ast}$ (\textit{bottom plot}). The observed ratios are shown with black triangles, while the corrected ones are plotted
  with red stars.}
\end{figure}

To study the spatial distribution of star-formation in the young stellar disc and compare it with the extent of stellar continuum optical emission, we plot in the upper 
plot of Fig.~\ref{fig:Rs_ratio_Mstar} the ratio of scalelengths in B band and corresponding to the H$\alpha$ line, both observed and intrinsic (corrected), as a function of
$M_{\ast}$. The observed and intrinsic (dust and inclination corrected) B band scalelengths were already determined in Paper I and II. For the observed
scalelength ratio, one would expect the measured scalelength of the stellar continuum optical emission to be larger than the $H\alpha$ line one. This is because dust effects,
which tend to artificially flatten the central parts of the disc surface brightness profiles (\citealt{Pas13a}) and therefore increase the measured disc scalelenghts due to
the more concentrated dust distribution towards the centre of the disc, are stronger at shorter wavelengths, as shown in \cite{Pas13a}. Therefore, one would measure a larger
disc scalelength in B band than for the H$\alpha$ disc, for the same galaxy. One can see that this is the case for about 2/3 of our galaxies, with a few values around 1.00
and a few outliers. On the other hand, comparing now the ratio of the intrinsic scalelengths, one needs to consider that the corrections due to dust effects are larger for
B band (as found in \citealt{Pas13a}). Thus, we would expect to obtain a scalelength ratio on average of $1.00$ (within errors) after applying the corrections, if there would
not be any inside-out growing of galaxy disc through star-formation. However, with the exception of a few outliers, for most of the galaxies this ratio is subunitary, 
meaning that on average, the distribution of star-formation in the disc is more extended than the optical emission one. It is also apparent that more massive galaxies have
a more extended H$\alpha$ disc than the B band optical one, compared with low mass ones, although this trend is rather weak. On average, we find for the corrected inverse ratio, 
$R_{s}^{d}(H\alpha)/R_{s}^{d}(B)$, a value of 1.10. This result is slightly lower but consistent with the $1.18\pm0.08$ value found by \cite{Math22} for their larger $z\sim0.5$
sample galaxies, considering also the fact that our sample galaxies are at much lower redshifts. Other studies of local Universe galaxies, such as those of \cite{Jam09} and 
\cite{Fos13}, have found the H$\alpha$ disc to have the same spatial extent as the optical stellar disc. On the other hand, other recent studies on higher redshift samples of
star-forming galaxies, such as those of \cite{Nel16} ($z\sim1$ sample, taken from the 3D-HST survey) and \cite{Wil20} ($z\sim1.7$ sample from the $KMOS^{3D}$, K-band Multi-Object
Spectrograph survey),  have revealed the star-formation disc to be slightly more extended than the stellar continuum one. This result is in agreement with our result and the one
of \cite{Math22}. However, a larger sample in our study would bring more clarity in this issue, at least for the local Universe galaxies. \\
In the bottom plot of Fig.~\ref{fig:Rs_ratio_Mstar}, the ratio of stellar mass surface densities for B band and H$\alpha$ line are plotted (in log scale), again as a function
of stellar mass. A slightly decreasing trend with stellar mass can be seen, more massive galaxies having a more compact stellar emission surface density than the star-formation
one, while for lower mass galaxies this difference is not so significant. This downward trend towards more massive galaxies has also been observed recently in \cite{Math22}
for their $z\sim0.5$ and $z\sim1$ galaxies. Overall, we find an average $\mu_{\ast}(H\alpha)/\mu_{\ast}(B)$ of 0.77 for our local sample, a value in very good agreement with
the result of $0.81\pm0.15$ found in \cite{Math22} for their $z\sim0.5$ sample.
 
\section{Discussion}\label{sec:discussion}

In this section, we are coming back to a few issues observed while analysing the main results, analyse the potential sources of systematic errors, and discuss the limitations of our
method.
\subsection{Potential sources of systematic errors}
A potential source of uncertainty is the calibration / conversion coefficient used to calculate SFR, as in Eq.~\ref{eq:SFR_obs}, which then produces effects in all the other
star-formation related quantities and the characteristics of the star-formation scaling relations. Various versions of this coefficient exist in the literature, depending on the
IMF and stellar evolution models considered, with values in the range $(4.4-7.9)\times10^{-42}$, as in \cite{Ken98}, \cite{Calz00}, \cite{Calz07}, \cite{Ken09}, \cite{Pes22},
\cite{Gime22}, etc. Due to this, a variation of up to 80$\%$ exists between various calibration coefficients, producing a significant uncertainty already at this point.
Star-formation rates can also be underestimated when using H$\alpha$ luminosities to derive it for galaxies with $L^{obs}(H\alpha)\leq2.5\times10^{39}$ (low SFR regions with 
$SFR\leq0.01M_{\bigodot}/yr$) as pointed out by \cite{Lee09}, even if the SFR are corrected for dust attenuation effects. However, this is not the case for any of our sample
galaxies.

Another important source of systematic errors affecting SFR determined from $L^{obs}(H\alpha)$, is the dust attenuation prescription used. As mentioned in 
Sec.~\ref{sec:corr}, the Balmer decrements or ratios between other hydrogen recombination lines (Paschen - Pa$\alpha$/H$\alpha$ , Pa$\beta$/H$\alpha$ or Brackett lines) have
been widely used in many other studies, with assumptions of a dust attenuation curve and/or a foreground dust screen aproximation. The Balmer decrement method however,
works mostly for normal galaxies, not in starbursts or dusty galaxies, and has large variations on small scales (\citealt{KenEv12}). Through the method proposed in this paper,
we circumvent these assumptions but we do introduce another potential source of uncertainty through the tailored \cite{Gro13} $\tau-\mu_{*}$ correlation and the relation 
between $\tau_{H\alpha}$ and $M_{dust}$ in Eq.~\ref{eq:Mdust_Halpha}, both relying on the \cite{Pop11} model, which assumes a fixed dust-star geometry and dust disc scale-height.
Moreover, a certain degree of uncertainty can arise from the choice of the dust model, its characteristics being encapsulated in the $K(H\alpha)$ constant, but this cannot
be avoided. Nevertheless, the relations in Eqs.~\ref{eq:Mdust_Halpha} and \ref{eq:Grootes1} have been calibrated on a representative large sample of low-redshift 
spiral galaxies and, our values for the attenuation of the emission line have been self-consistently derived with this method. Moreover, the sample analysed here is composed
mostly by normal spiral and lenticular galaxies, not galaxies with a more peculiar geometry or starburst galaxies.\\
Therefore, we believe that the introduced systematic errors related to the dust attenuation treatment used here, have a less significant impact on the main results (SFR
and related quantities, the galaxy scaling relations characteristics) than the ones introduced by the other mentioned dust attenuation approaches, allowing more 
accurate and less biased results and conclusions.\\
A third source of uncertainties, this time only for the characteristics of the scaling relations analysed, can come from the choice of the regression routine used, 
which can produce a better representation of the fitted data, it may affect significantly the slope and the zero-point of the relation, and further, its $\sigma$
parameter. Observational uncertainties can also determine a certain degree of intrinsic scatter in the parameters associated with most scaling relations, this effect
being difficult to assess, as pointed out by \cite{Sto21}, and not taken into account in most studies. For our small sample, we find that the ordinary least-squares
(OLS) routine used here (where necessary) outputs a best-fit that reproduces the trends seen in the data, with a high degree of accuracy. However, we do recognise that
more complex routines, such as orthogonal distance regression (ODR), linear bisector regression algorithms (BCES, \citealt{AkB96}) or others, capable of taking into
account covariant uncertainties, can and should be used for larger scale studies, as needed.

In Section~\ref{sec:results}, we compared our results for the main characteristics of the scaling relations and the trends observed with results from other similar studies.
While most of the results were consistent within errors with the compared studies, there were also some noticeable differences pertaining to the lack of a correlation in
some cases, or inconclusive / different trends for other relations. We also compared our results with those found in studies done on samples of low redshift local galaxies,
similar to our sample, or at the closest redshift possible, to make the comparison more meaningful. As mentioned earlier, inconclusive trends (e.g. $M_{dust}/M_{star}-sSFR$)
can be attributed to the low statistics in this work. Results in apparent opposition with other studies (e.g. $\tau_{H_{\alpha}}^{f}-sSFR$, $\tau_{H_{\alpha}}^{f}-\Sigma_{SFR}$)
or in agreement only with some ($M_{dust}/M_{star}-sSFR$) may be explained through an inadequate treatment of the biases introduced by dust and inclination effects on the
measured parameters in the respective studies, or other systematic biases.

\subsection{Limitations of the method and range of applicability}

The limitations of the proposed method (and its succesion of steps) and its range of applicability are tightly connected to the range of applicability of the tailored
Eq.~\ref{eq:Grootes1} \& Eq.~\ref{eq:Mdust_Halpha}, and also of the numerical corrections for dust and inclination (projection) effects used here, from \cite{Pas13a,Pas13b}.
The latter are destined to be applied for normal low to intermediate redshift spiral, elliptical and lenticular galaxies, but not for the more irregular geometries, dwarf,
disturbed or peculiar shape galaxies, as they were derived using simulated images produced by radiative transfer calculations, with a typical fixed star-dust geometry, 
considered in \cite{Pop11}. Eq.~\ref{eq:Grootes1} and Eq.~\ref{eq:Mdust_Halpha} also depend on the range of applicability of the large-scale geometry of the exponential 
dust disks as calibrated in the \cite{Pop11} model to the range of galaxy types, morphologies (same as for the dust and inclination effects numerical corrections) and stellar
mass surface densities, $8.0 \leq log(\mu_{\ast}) \leq 11.0$ (therefore intermediate mass galaxies), one has to analyse with this method.\\
The method presented in this work can be used for larger scale studies of star-formation and ISM evolution, for which H$\alpha$ imaging is or will be available
(e.g. J-PAS, J-PLUS, S-PLUS surveys, VLT-MUSE, and potentially others). The main advantages are: a) only H$\alpha$ fluxes / luminosities are needed (compared with other methods where
MIR, TIR or FUV are required, or even other hydrogen recombination line fluxes); b) it is easy and straightforward to use based on Eqs.~\ref{eq:Mdust_Halpha}, \ref{eq:Grootes1},
\ref{eq:L_Halpha_corr} and the numerical corrections from \cite{Pas13a,Pas13b}; c) no Balmer decrements needed or assumption of a certain attenuation curve; d) it can be
applied for normal low to intermediate redshift spiral, elliptical and lenticular galaxies; e) more instantaneous SFRs can be derived through it and therefore, it can be
also used as a means to compare with coresponding SFR values derived through other methods, which determine less recent galaxy SFR values.

\section{Summary and conclusions}\label{sec:conclusions}

In this paper we have presented a detailed analysis of dust/ISM and star-formation scaling relations of a small representative sample of typical nearby spiral and 
lenticular galaxies taken from the SINGS/KINGFISH survey. This was done with the purpose of: i) investigating the changes induced by dust and inclination effects in the 
characteristics of these relations (slope, zero-point, scatter and correlation coefficients) and in the SFR values; ii) understanding which relations are fundamental and which
are derived, or are a consequence of others; iii) verifying which of the derived specific parameters are actually correlated and why (besides the already established relations)
and which relations are tighter (reduced degree of scatter) than others; iv) test the proposed method on a well studied sample of galaxies and check its accuracy and consistency
compared with other similar studies in the literature.\\
For this purpose, the $H\alpha$ optical emission line flux was chosen as a SFR tracer and the H$\alpha$ line images were used and analysed in order to derive the integrated
fluxes and luminosities, needed further for determination of the star-formation rate of each galaxy. The succesion of steps as in Papers I and II was followed
for the photometry, structural analysis and the calculation of corrected parameters involved in the analysed scaling relations. We used again the empirical relation found
by \cite{Gro13}, slightly modified for the H$\alpha$ line wavelength to determine the central face-on dust opacity, $\tau_{H\alpha}$, needed when applying the corrections for dust
attenuation effects.
We derived the SFR using the unattenuated H$\alpha$ luminosities to obtain more accurate and instantaneous star-formation rate values than would be derived through other
methods. \\
The new self-consistent method proposed here to determine the corrected H$\alpha$ luminosities and star-formation rates, does not require the consideration of a
dust attenuation curve and the use of Balmer decrements or other hydrogen recombination lines to estimate the dust attenuation, as in many other similar studies in the literature.
It offers a more accurate and consistent treatment of the dust effects, which hampers the accurate measurements of SFRs and related quantities, compared with other
studies which use completely different methods, and therefore produces less biased values for these and the characteristic parameters of the specific scaling relations.
For most of the corrected relations we investigated the degree of correlation between the parameters, calculated the scatter of these relations and analysed the 
implications of the main results for star-formation and galaxy evolution. Some of the analysed scaling relations are already known (e.g. SFMS, rSFMS, $M_{dust}-M_{HI}$
$M_{dust}/M_{HI}-M_{\ast}$), while others have been less studied previously (e.g. SFR or sSFR vs $\mu_{\ast}$; dust opacity vs SFR, sSFR, $M_{\ast}$, and $\Sigma_{SFR}$).
Our main results are:
\begin{itemize}
 \item the corrected SFMS and $sSFR-M_{\ast}$ trends and characteristic parameters obtained are consistent within errors with those found in similar studies, with $sSFR-M_{\ast}$ correlation
 being weaker than the SFMS one;
 \item no correlation between corrected SFR and $\mu_{\ast}$ is observed, while the $SFR-\mu_{\ast}$ slope is shallower than the SFMS one and less tight;
 \item the $M_{dust}-SFR$ correlation is recovered, with a higher degree of scatter; we appreciate that this relation exist as a consequence of the tighter 
 $M_{dust}-M_{\ast}$ and $SFR-M_{\ast}$ relations, with better statistics needed for a more clear answer;
 \item an expected apparent downward trend of dust-to-stellar mass ratios with SFR was observed, while no conclusive evolution between $M_{dust}/M_{\ast}$ and sSFR was found;
 \item the resolved SFMS is recovered with an almost linear slope of $1.03\pm0.18$ (in log scale) within range of those obtained in a more detailed way, with a high correlation
 coefficient (0.79, comparable with the global SFMS one), but with a larger scatter; 
 \item the H$\alpha$ face-on optical depth is found to increase with SFR and $M_{\star}$, a consequence of the $M_{dust}-SFR$ and Eq.~\ref{eq:Mdust_Halpha}, but also with
 $\Sigma_{SFR}$, in agreement with other works; no dependence of $\tau_{H\alpha}$ on corrected sSFR was found 
 \item we confirm the increase of dust-to-gas ratio (HI) towards more massive galaxies, but not with $\mu_{\ast}$, while the characteristic parameters are consistent within errors with those found by other authors; the average $M_{dust}/M_{HI}$ of $0.63\%$ for our sample is consistent with $\simeq1\%$ the dust mass fraction in the ISM of
 normal galaxies; the strong correlation between $M_{dust}$ and $M_{HI}$ is also confirmed even for this small sample;
 \item we compared the B band optical disc scalength and H$\alpha$ line (star-forming) disc one and found that on average, the latter is more extended than the stellar continuum
 optical one (with a ratio of 1.10), this extent being larger for more massive galaxies; similarly, more massive galaxies have a more compact stellar emission surface density
 than the star-formation one, this behaviour being less apparent for lower mass galaxies; we find an average $\mu_{\ast}(H\alpha)/\mu_{\ast}(B)$ of 0.77 for our local sample.
\end{itemize}
We compared the main results and the trends seen for the analysed scaling relations and found most of these to be consistent with other results in the literature.
While this study has been done on a small sample but representative galaxies in the local Universe, to test the consistency and accuracy of the proposed method,
we advocate that this method can be used in future larger scale studies of star-formation and ISM evolution for low to mid-redshift galaxies. This work underlines the
importance of having accurate, unbiased scaling relations in models of ISM evolution and star-formation.

\section*{Acknowledgements}
The author would like to thank the referee for a very careful reading of the manuscript and for the useful suggestions which improved the quality and clarity of this paper.\\
This research made use of the NASA/IPAC Extragalactic Database (NED), which is operated by the Jet Propulsion Laboratory, California Institute of Technology, under contract
with the National Aeronautics and Space Administration.\\
This research was supported by Romanian Ministry of Research, Innovation and Digitalization under Romanian National Core Programs LAPLAS VI/2019 and LAPLAS VII - contract
no. 30N/2023.
\section*{Data Availability}
The data underlying this article are available in the article.\\

\appendix
\section{Reevaluation of the $\tau_{B}-\mu_{\ast}$ relation for the $H\alpha$ line}
To evaluate the changes needed for Eq.~\ref{eq:Grootes} to be valid at the $H\alpha$ line wavelength, we came back to the detailed derivation of dust opacity in the paper
of \cite{Gro13}, namely Eqs. (1), (2) in section 2 and more important, the suite of relations (A1-A9), thoroughly described in their Appendix A. Following the equations,
the main change in the expresion of $\tau_{\lambda}^{f}$ would be in the value of the factor $A$ (See Eqs. (A8) \& (A9) in the same paper), empirically calibrated to the
value of 6.939$\times10^{-13}$ $arcsec^{2} J/Jy/s/Hz/m^{2}/sr$  based on \cite{Pop11} model. 
This is because a factor of $\gamma^{2}$ was introduced in the expression of $A$ by the authors to convert the disc scalelength of the disk in B band to the corresponding 
one in $r$ band (suitable for their analysis), as $\gamma=R_{s,d}(B)/R_{s,d}(r)$. Its value was derived considering the fixed geometry of the \cite{Pop11} model, where the
intrinsic disc scalength of the disc decreases with wavelength ($R_{s,d}(r)$ being smaller than $R_{s,d}(B)$), with this ratio being $\gamma=1.067$ at $r$ band
wavelength of $6600\AA{}$.\\
Leaving the other reference values unchanged, we recalculated the value of factor $A$ for our case by multipliying the value derived in \cite{Gro13} with $\gamma^{2}$ and
then dividing it with another $\gamma^{2}_{H\alpha}$ term, to account for the conversion of disc scalelenghts $R_{s,d}(B) \Rightarrow R_{s,d}(H\alpha)$. To derive this new
value we again considered the same geometry of the \cite{Pop11} model and interpolated their values at the $\lambda_{H\alpha}=6563\AA{}$ wavelength. We derived a 
$\gamma_{H\alpha}=1.074$ value and a new factor $A$ with a corresponding value of 6.852$\times10^{-13}$ $arcsec^{2} J/Jy/s/Hz/m^{2}/sr$. As a
result, the changes induced in the slope and intercept of the correlation will be within the standard deviations derived by \cite{Gro13}, of $\pm0.11$ and $\pm0.8$. \\
We can therefore rewrite Eq.~\ref{eq:Grootes} as
\begin{eqnarray}\label{eq:Grootes1a}
\log(\tau_{H\alpha}^{f})=1.12(\pm0.11)\cdot\log(\mu_{*,H\alpha}/M_{\odot}kpc^{-2})-8.6(\pm0.8) ,
\end{eqnarray}
with $\mu_{*,H\alpha}$ being the stellar mass surface density (derived using the scalelength of the H$\alpha$ disc obtained through the bulge-disc decomposition).\\

\bsp	
\label{lastpage}
\end{document}